\documentclass[a4paper,11pt]{article}

\pdfoutput=1 

\usepackage{jcappub} 
                     
\usepackage{mathtools, cases}
\usepackage[T1]{fontenc} 
\usepackage{float}
\usepackage{xcolor}
\usepackage{caption}

\newcommand{\bk}[1]{\boldsymbol{k}_{#1}}
\newcommand{\bq}[1]{\boldsymbol{q}_{#1}}
\newcommand{\bu}[1]{\boldsymbol{u}_{#1}}
\newcommand{\bx}[1]{\boldsymbol{x}_{#1}}

\definecolor{tabblue}{HTML}{1f77b4}
\definecolor{tabgreen}{HTML}{2ca02c}
\definecolor{tabred}{HTML}{d62728}

\title{\boldmath Improving constraints on primordial non-Gaussianity using neural network based reconstruction}

\author[a,b]{Thomas Fl\"oss,}
\author[a]{P. Daniel Meerburg}

\affiliation[a]{Van Swinderen Institute, University of Groningen, Nijenborgh 4, 9747 AG Groningen, The Netherlands}
\affiliation[b]{Kapteyn Astronomical Institute, University of Groningen, P.O.Box 800, 9700 AV Groningen, The Netherlands}

\emailAdd{tsfloss@gmail.com}

\abstract{We study the use of U-Nets in reconstructing the linear dark matter density field and its consequences for constraining cosmological parameters, in particular primordial non-Gaussianity. Our network is able to reconstruct the initial conditions of redshift $z=0$ density fields from N-body simulations with $90\%$ accuracy out to $k \leq 0.4$ h/Mpc, competitive with state-of-the-art reconstruction algorithms at a fraction of the computational cost. We study the information content of the reconstructed $z=0$ density field with a Fisher analysis using the \texttt{QUIJOTE} simulation suite, including non-Gaussian initial conditions. Combining the pre- and post-reconstructed power spectrum and bispectrum data up to $k_{\rm max} = 0.52$ h/Mpc, we find significant improvements in all parameters. Most notably, we find a factor $3.65$ (local), $3.54$ (equilateral), and $2.90$ (orthogonal) improvement on the marginalized errors of $f_{\rm NL}$ as compared to only using the pre-reconstructed data. We show that these improvements can be attributed to a combination of reduced data covariance and parameter degeneracy. The results constitute an important step towards a more optimal inference of primordial non-Gaussianity from non-linear scales.}

\begin{document}
\maketitle
\newpage
\section{Introduction}
Primordial non-Gaussianity (pnG) provides an important window into the early universe and in particular the inflationary epoch \cite{Achucarro:2022qrl}. It provides insight into the physics at play during this epoch, from the fields responsible for the accelerated expansion and their interactions \cite{Meerburg:2019qqi} to that of auxiliary particles present during the epoch \cite{Arkani-Hamed:2015bza,Lee:2016vti}. So far, pnG has been most precisely constrained using the anisotropies of the cosmic microwave background (CMB) \cite{Ferreira:1998kt,Komatsu:2001wu,WMAP:2003xez,Creminelli:2005hu,Planck:2015zfm,Planck:2019kim}. These anisotropies, being linearly related to the small density fluctuations sourced by inflation, provide an ideal probe for studying pnG. Observations by the Planck satellite have therefore led to the most stringent constraints on pnG to date which are consistent with Gaussian initial conditions  \cite{Planck:2019kim}. Many models of inflation predict primordial non-Gaussianity below current upper limits. Although upcoming experiments aim to improve constraints on pnG significantly \cite{SimonsObservatory:2018koc,CMB-S4:2016ple}, the CMB is ultimately limited in the number of observable modes, due to its 2D nature and diffusion (Silk) damping at small scales \cite{Kalaja:2020mkq}. Advancing our understanding of the primordial universe will therefore eventually depend on our ability to constrain pnGs with additional probes.\\

The small anisotropies observed in the CMB evolve into the large-scale structure (LSS) of the universe. The three-dimensional comoving volume of the universe, spanning all the way from the Dark Ages until now, contains exponentially more modes than the CMB, making it a powerful probe for pnGs (see e.g. Refs.~\cite{Munoz:2015eqa,Meerburg:2016zdz,Karagiannis:2020dpq,Floss:2022grj,Cabass:2022epm} for forecasts). The underlying dark matter density distribution can be studied using biased tracers such as galaxies and atomic and molecular spectral lines \cite{Pritchard:2011xb,Kovetz:2017agg}. Except on the largest scales, contrary to the anisotropies in the CMB, anisotropies in the LSS are intrinsically non-linear due to the non-linear nature of gravity, the formation of virialized bound structures, and astrophysical processes. This naturally results in a highly non-Gaussian density distribution. In order to improve over constraints from the CMB, we need to be able to extract information from within this non-linear regime, requiring an accurate model of the tracer. Over the past years, major progress has been made to address this theoretical challenge \cite{Bernardeau:2001qr,Taruya:2012ut,Carrasco:2012cv}. Although so far LSS surveys have put constraints on pnG that are not yet competitive with CMB \cite{Slosar:2008hx,Ross:2012sx,Leistedt:2014zqa,Ho:2013lda,Castorina:2019wmr,Mueller:2021tqa,Cabass:2022wjy,DAmico:2022gki,Cabass:2022ymb}, these intermediate results provide confidence and motivation to further pursue this avenue. Besides the complication of modeling the tracer field, recent work has emphasized the additional complication of mode coupling or \emph{non-Gaussian covariance} \cite{Chan:2016ehg}, especially in the context of primordial non-Gaussianity \cite{Biagetti:2021tua,Coulton:2022qbc,Floss:2022wkq,Goldstein:2022hgr}. Non-linear evolution couples modes of different wavelengths, resulting in a reduction of the amount of unique information contained in each mode. As a result, modeling modes deeper into the non-linear regime yields diminishing returns in terms of information, saturating parameter constraints. In order to improve constraints, additional work needs to be done to undo this mode coupling. Put in another way, non-linear evolution spreads the information content of the initial conditions into all N-point statistics. Reconstructing the initial (linear) field thus brings information from higher-order correlation functions back to the power spectrum and bispectrum. This idea has been especially well tested in the context of reconstructing the BAO peak \cite{Eisenstein:2006nk}. It has also been shown that reconstruction indeed results in improved parameter constraints \cite{Wang:2022nlx,Shirasaki:2020vkk}. From this perspective, in this work we will further investigate how much of the information on cosmological parameters that is lost due to non-linear gravitational evolution can be recovered by reconstructing the initial conditions at the field level.\\

Given the non-linear nature of gravitational evolution, it naturally lends itself to machine learning applications. Previously, machine learning has been applied to solve the \emph{forward} problem of emulating the outcome of N-body simulations given some initial condition \cite{He:2018ggn,Jamieson:2022lqc}. Moreover, it was shown that neural networks trained to emulate N-body simulations learn general properties of gravitational evolution \cite{Jamieson:2022daw}. Previous attempts at reconstructing the initial conditions using machine learning have shown promising results \cite{Jindal:2023qew,Shallue:2022mhf}.\\

Our work consists of two parts. First, we develop a neural network based approach to reconstructing the linear initial conditions of the late-time dark matter density field that is competitive with state-of-the-art iterative reconstruction methods (i.e.~Ref.~\cite{Schmittfull:2017uhh}) at a fraction of the computational cost. Secondly, applying our reconstruction methodology to the \texttt{QUIJOTE} simulation suite \cite{Villaescusa-Navarro:2019bje,Coulton:2022qbc}, we are able to determine the improvement of marginalized parameter constraints when using pre- and post-reconstructed power spectrum and bispectrum measurements. As we will see, our reconstruction method reduces mode coupling (covariance) as well as degeneracy between cosmological parameters. This results in significant improvements of parameter constraints, in particular for primordial bispectra (up to a factor of $3.65$). Such improvements are hard to realize by including more non-linear modes, both because of the complications in modeling them (however, see e.g. \cite{Goldstein:2022hgr,Biagetti:2022ckz} for ways around this issue) as well as the aforementioned saturation due to mode coupling.\\

This paper is organized as follows. In Section \ref{sec:NG} we discuss the non-Gaussianity of the dark matter density field and its implications for cosmological parameter inference. Section \ref{sec:REC} discusses reconstruction algorithms followed by an extensive discussion of our neural network based reconstruction method. In Section \ref{sec:INFO} we study the implications of our reconstruction methodology on parameter constraints. We conclude in Section \ref{sec:conclusion}. 

\section{Non-Gaussianity of the dark matter density field}
\label{sec:NG}

\subsection{Primordial non-Gaussianity}
In this work, we will focus on the primordial three-point correlation function of the primordial potential $\Phi(\bx{})$, or its Fourier space equivalent, the primordial bispectrum:
\begin{align}
    \langle \Phi_{\bk{1}} \Phi_{\bk{2}} \Phi_{\bk{3}}\rangle = \left(2\pi\right)^3 \delta_{\rm D}^{(3)}(\bk{1} + \bk{2} + \bk{3})B_\Phi(\bk{1},\bk{2},\bk{3})
\end{align}
where spatial homogeneity forces the momentum space triangles to be closed. Different inflationary scenarios source bispectra with distinct functional dependence on the triangle configuration, often referred to as the \emph{shape} of the bispectrum \cite{Babich:2004gb}. In order to look for primordial non-Gaussianity in data, bispectrum templates have been developed that cover general features of classes of inflationary theories parameterized only by their overall amplitude $f_{\rm NL}$. We will study three of these templates. The first of them, known as the \emph{local} shape is given by
\begin{align}
    B^{\rm local}_\Phi(k_1,k_2,k_3) = 2 f^{\rm local}_{\rm NL}\left(P_\Phi(k_1)P_\Phi(k_2) + 2\rm{\,perms.}\right),
\end{align}
which peaks in the \emph{squeezed limit}, i.e. when  $k_1 \ll k_2 \sim k_3$. This type of non-Gaussianity is generically suppressed in inflationary models with only a single field driving the accelerated expansion, such as slow-roll inflation \cite{Maldacena:2002vr,Creminelli:2004yq}. Detection of a large bispectrum of this kind would be a strong hint toward inflationary scenarios with multiple fields. Cubic self-interactions of the inflaton, e.g. $(\partial\Phi)^3$, give rise to a bispectrum that is largest for equilateral triangle configurations, i.e. when $k_1 = k_2 = k_3$, captured by the template:
\begin{align}
    B_\Phi^{\rm equil}(k_1,k_2,k_3) = 6 f_{\rm NL}^{\rm equil}\Big(- P_\Phi(k_1)P_\Phi(k_2) + 2 \rm{\, perms.} -2 \left(P_\Phi(k_1)P_\Phi(k_2)P_\Phi(k_3)\right)^{2/3} \nonumber \\  + P_\Phi(k_1)^{1/3} P_\Phi(k_2)^{2/3}P_\Phi(k_3) + 5 \rm{\,perms.}\Big).
\end{align}
These types of interactions naturally arise in the Effective Field Theory (EFT) of Inflation \cite{Cheung:2007st}. Another prediction of this EFT is the orthogonal shape, covered by the template \cite{Senatore:2009gt}:
\begin{align}
    B_\Phi^{\rm orth}(k_1,k_2,k_3) &= 6 f_{\rm NL}^{\rm orth}\left(\left(1+p\right) \frac{\Delta_{123}}{k_1^3k_2^3k_3^3} - p \frac{\Gamma^3_{123}}{k_1^4 k_2^4 k_3^4}\right), \\
    \Delta_{123} &= (k_T - 2k_1)(k_T - 2k_2)(k_T - 2k_3), \\
    \Gamma_{123} &= \frac{2}{3}(k_1 k_2 + k_2 k_3 + k_3 k_1)-\frac{1}{3} (k_1^2 + k_2^2 + k_3^2), \\
    p & = \frac{27}{-21 + \tfrac{743}{7(20\pi^2 - 193)}}
\end{align}

The imprint of primordial non-Gaussianity on the large-scale structure of the universe is an active field of research. When studied in perturbation theory, the lowest order contribution of the primordial bispectrum to the dark matter density statistics is through the linearly evolved bispectrum:
\begin{align}
\label{eq:treeprimB}
    B_\delta(k_1,k_2,k_3) = \mathcal{M}(k_1)\mathcal{M}(k_2)\mathcal{M}(k_3)B_{\Phi}(k_1,k_2,k_3),
\end{align}
where $\mathcal{M}(k)$ is the linear transfer function for dark matter fluctuations. Additionally, there are higher-order non-linear corrections to the dark matter power spectrum and bispectrum that correspond to loop corrections in the perturbation theory.

\subsection{Gravitational non-Gaussianity}
Although the universe at the time of last-scattering appears Gaussian, the late-time universe does not. Limiting ourselves to the dark matter density field underlying the large-scale structure of the universe, this non-Gaussianity is a consequence only of the non-linear gravitational evolution. Any small amount of non-Gaussianity in the initial conditions is therefore obscured by the strong gravitational non-Gaussianity. To put reliable constraints on pnGs from LSS requires an accurate model of the gravitationally induced non-Gaussianity. In addition, any uncertainties in the parameters of the model need to be marginalized over to avoid bias in the primordial parameter $f_{\rm NL}$ that could result in a false detection. \\

The highly non-linear nature of gravity makes it challenging to model using standard perturbation theory (SPT, \cite{Bernardeau:2001qr}). To improve predictability, modifications to SPT, such as regPT \cite{Taruya:2012ut} have been proposed. The Effective Field Theory of Large Scale Structure (EFTofLSS) \cite{Carrasco:2012cv} is a self-consistent expansion of the fluctuations, and limitations and corrections are well understood. EFTofLSS is currently considered the most reliable analytical perturbation expansion and it can accurately model the dark matter power spectrum and bispectrum out to scales $k\leq 0.6$ h/Mpc at the cost of having to marginalize over additional parameters \cite{Carrasco:2013mua, Angulo:2014tfa}.\\

To push beyond these scales accessible with analytical methods, we rely on numerical N-body simulations of the dark matter field, which evolve an initial distribution of dark matter particles in a periodic box under their gravitational interaction. In principle, running many of these simulations and measuring their power spectrum and bispectrum provide an accurate representation to deep within the non-linear regime.  The initial conditions can be generated assuming varying cosmological parameters, which allows studying the imprint of these parameters on the late-time statistics of the dark matter field.

\subsection{Non-Gaussian covariance}
Besides swamping any primordial signal, gravitational evolution can couple modes of different wavelengths. In the absence of this mode coupling,  every Fourier mode in the density field is its own Gaussian random variable, independent of all the other modes. This remains a good approximation when there is only a weak coupling, such as a weak primordial bispectrum (e.g. in the CMB, however, see also \cite{Coulton:2019odk}). In the presence of strong mode coupling, the information contained in the initial density field is scrambled across N-point statistics at late times. Hence, even if we are able to accurately model the power spectrum and bispectrum, the amount of information available from just the power spectrum and bispectrum is limited by this mode coupling. When inferring cosmological parameters from data, this mode coupling is captured by off-diagonal components of the covariance matrix that enters the likelihood function. These entries to the covariance matrix (together with higher N-point contributions to the diagonal) are referred to as non-Gaussian covariance. The covariance matrix can in principle be modeled analogous to the power spectrum and bispectrum signals itself, although the computation involves higher-order correlation functions (see e.g. \cite{Chan:2016ehg, Biagetti:2021tua}), which we can model analytically only on large scales, or low $k$. Instead, using N-body simulations, the covariance matrix can be determined numerically by taking the measurements of many simulation realizations. \\

The covariance matrix plays a vital part in a Fisher analysis that forecasts the expected error bars on cosmological parameters for a particular survey (e.g. galaxies or intensity mapping). Forecasts derived by analytical calculations often assume only the Gaussian (diagonal) contribution. In Ref.~\cite{Chan:2016ehg} it was shown that this assumption is unwarranted as non-Gaussian covariance of the dark matter and halo power spectrum and bispectrum significantly reduce the signal-to-noise ratio already in what is usually considered to be the linear regime. Recently, Refs.~\cite{Biagetti:2021tua} showed that the bispectrum covariance is dominated by squeezed triangle configurations, significantly affecting constraints on local primordial non-Gaussianity. This finding was extended to other shapes of non-Gaussianity using the recently released \texttt{QUIJOTE} simulations with non-Gaussian initial conditions \cite{Coulton:2022qbc}. Finally, it was shown that the local shape remains significantly affected even at higher redshifts, where the field is more linear but where we also expect to access more squeezed triangle configurations \cite{Floss:2022wkq}. In order to improve constraints on pnGs from the power spectrum and bispectrum, we need to find ways to reduce the effects of mode coupling. This is the topic of the rest of this paper.

\section{Reconstructing the initial conditions}
\label{sec:REC}
Since the coupling of modes is a consequence of the non-linear gravitational evolution of the density field, we expect to be able to access more information by reversing the gravitational collapse. This idea of \emph{reconstruction} has been studied extensively in the context of BAOs, where it is used to recover the BAO peak in the two-point correlation function that is otherwise washed out due to the gravitational collapse \cite{Eisenstein:2006nk}. It has been shown that this BAO reconstruction method can aid in reducing non-Gaussian covariance and improving parameter constraints \cite{Wang:2022nlx}, especially in the context of primordial non-Gaussianity \cite{Shirasaki:2020vkk}. The latter work only considered the tree-level primordial contribution \eqref{eq:treeprimB} to the bispectrum, whereas in the non-linear regime, we expect additional loop contributions to both the power spectrum and bispectrum. Since reconstruction affects the loop contributions (and thereby the information contained in the power spectrum and bispectrum), any assessment of the improvement due to reconstruction should include the full non-linear power spectrum and bispectrum. Furthermore, we expect to be able to improve constraints using more sophisticated reconstruction methods than the standard reconstruction algorithm.

\subsection{Reconstruction Algorithms}
\label{sec:recalgo}
Although a full review of existing reconstruction algorithms is beyond the scope of this work (see e.g. \cite{Schmittfull:2015mja}), we briefly summarize here several methods and their performance for completeness and comparison. Standard reconstruction (or Lagrangian-Growth-Shift), as originally developed for reconstructing the BAO feature in the matter power spectrum, relies on approximating the linear displacement field in order to move structure back to their initial positions \cite{Eisenstein:2006nk}. When performed on a redshift $z=0$ N-body simulation snapshot, the reconstructed density field correlates up to $\sim 37\%$ with the initial conditions at $k = 0.4$ h/Mpc (see Figure \ref{fig:crosscorr_real}). Since this original proposal, more sophisticated methods have been developed both at the level of the object (e.g. halo) catalog (Lagrangian space) \cite{Schmittfull:2017uhh} as well as the density field directly (Eulerian space) \cite{Schmittfull:2015mja}. The iterative reconstruction algorithm of Ref.~\cite{Schmittfull:2017uhh} can be considered the state-of-the-art non-machine learning approach, with a reconstruction cross-correlation of $\sim 90\%$ out to scales $k \leq 0.4$ h/Mpc at redshift $z=0$.

\subsection{Reconstruction using U-Nets}
In this work, we will focus on the application of neural networks in the context of reconstruction. Previous work has shown that a combination of standard reconstruction and convolutional neural networks is able to reconstruct the initial density field significantly better than standard reconstruction alone \cite{Shallue:2022mhf}. Instead, we defer the entire reconstruction to a neural network applied directly to the density field. In particular, we will employ a U-Net, which was originally developed for medical image segmentation \cite{ronneberger2015u}. Recently, there have been many successful applications of U-Nets in cosmology and astrophysics addressing problems that involve the coupling of different length scales (see e.g. \cite{Makinen:2020gvh,Gagnon-Hartman:2021erd}). Ordinary convolutional neural networks have a relatively small receptive field (depending on the size of the filter and the number of convolutions), making them local models. U-Nets incorporate down-sampling, effectively resulting in a larger receptive field, allowing them to also access non-local (large-scale) information, which is essential for our application. \\

\subsubsection{Network architecture}
\begin{figure}
    \centering
    \includegraphics{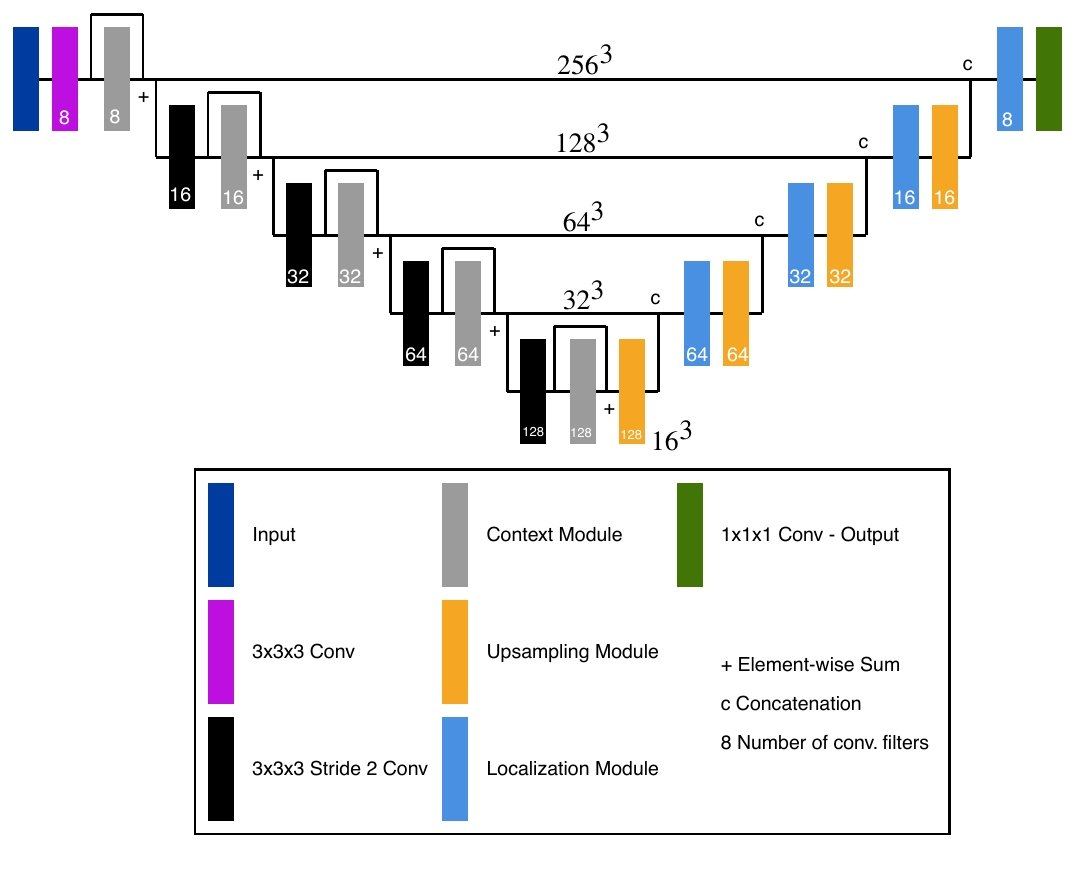}
    \caption{Diagrammatic representation of the neural network architecture used in this paper. The network is based on the one presented in Ref.~\cite{Gagnon-Hartman:2021erd}.}
    \label{fig:architecture}
\end{figure}

The architecture used in this work resembles the U-Net developed in Ref.~\cite{Gagnon-Hartman:2021erd}. The network takes a density field on a $256^3$ grid. Inspired by grid-based perturbation theory \cite{Taruya:2018jtk}, we compute the following velocity fields on-the-fly:
\begin{align}
    \bu{}(\bx{}) &= \int d^3\bk{} \; \frac{-i\bk{}}{k^2} \delta_m(\bk{}) e^{i \bk{}\cdot \bx{}}, \\
    \partial_i (\bu{})_j(\bx{}) &= \int d^3 \bk{} \; \frac{k_i k_j}{k^2} \delta_m(\bk{}) e^{i \bk{}\cdot \bx{}},
\end{align}
yielding six additional input fields. Subsequently, we apply an ordinary convolutional layer with a kernel of size $3^3$, followed by a \emph{context block}, consisting of two additional convolutional layers of size $3^3$, after which we add the output of the initial convolution to the output of the context block, this is known as a \emph{residual connection}. We repeat this, only now employing a stride of two in the initial convolutional layer, effectively down-sampling the feature maps by a factor of two in every spatial dimension. This is then repeated three more times until the feature maps have a size of $16^3$ cells. We then upsample the feature maps by a factor of two and concatenate the output of the residual connections in the down-sampling part of the network, forming what is known as \emph{skip connections}. The concatenated feature maps are fed into a \emph{localization block}, consisting of an ordinary convolutional layer of size $3^3$ followed by a convolutional layer of size $1^3$. This is repeated until we have reached the input size of $256^3$, after which we perform one more convolution of size $1^3$, in order to output a single feature map, representing the reconstructed density field. All but the last convolutional layer consists of the convolution, \emph{instance normalization}, and a \emph{Leaky ReLU} activation. The complete architecture is visualized in Figure \ref{fig:architecture}. The network has roughly 2 million parameters implemented using Keras/Tensorflow and is publicly available including trained weights.\footnote{\url{https://github.com/tsfloss/URecon}}

\subsubsection{Training}
The training dataset consists of 48 pairs of initial ($z=127$) and final ($z=0$) density fields of the fiducial cosmology \texttt{QUIJOTE} simulations, generated from the dark matter particle snapshots using a Piecewise Cubic Spline (PCS) mass assignment scheme to a grid of $256^3$ cells.\footnote{This might seem like a small dataset, but every simulation constitutes $256^3$ data points that need to be mapped to equally as many outputs. Training can be done with an even smaller dataset than the one considered here.} The aim of our network is to minimize the loss function that is the mean squared error (MSE) between reconstructed and initial density fields:
\begin{align}
    L_{\rm MSE} = \frac{1}{N_{\rm sims}}\sum_{i}^{N_{\rm sims}} \frac{1}{N_{\rm cells}}\sum_{\bx{}}\left(\delta^{(i)}_{\rm recon}(\bx{}) - \delta^{(i)}_{\rm init}(\bx{}) \right)^2,
\end{align}
where $\delta^{(i)}_{\rm recon}$ and $\delta^{(i)}_{\rm init}$ are the reconstructed and initial ($z=127$) density field of simulation $i$. It can be shown that the Fourier transform of this loss function is the mean squared error between different wavelength modes of the density field \cite{Shallue:2022mhf}. \\

We will eventually be interested in determining the information content of cosmological parameters of the power spectrum and bispectrum of the reconstructed density field. The FFT-based estimator used to compute the bispectrum can only be used up to a scale of $k_{\rm max} = \tfrac{2}{3}k_{\rm Nyquist}\approx 0.53$ h/Mpc \cite{Sefusatti:2015aex}. We have also explicitly checked that the power spectrum and bispectrum measurements up to this scale do not suffer from any aliasing effects, by comparing them to measurements made at a higher resolution grid of $1024^3$ cells. Since we are only interested in a sub-sample of the modes present in our density field, we limit our target initial density fields already to this same $k_{\rm max}$ using an isotropic cutoff in Fourier space. This has the advantage that the network does not have to reconstruct modes that we will not use in our final analysis, speeding up the training process. We train the network using 4 Nvidia A100 40GB GPUs until we see no more significant improvement in the validation loss, taking a bit under 4 hours.\\

\subsubsection{Network performance}
\begin{figure}
    \centering
    \includegraphics[scale=.61]{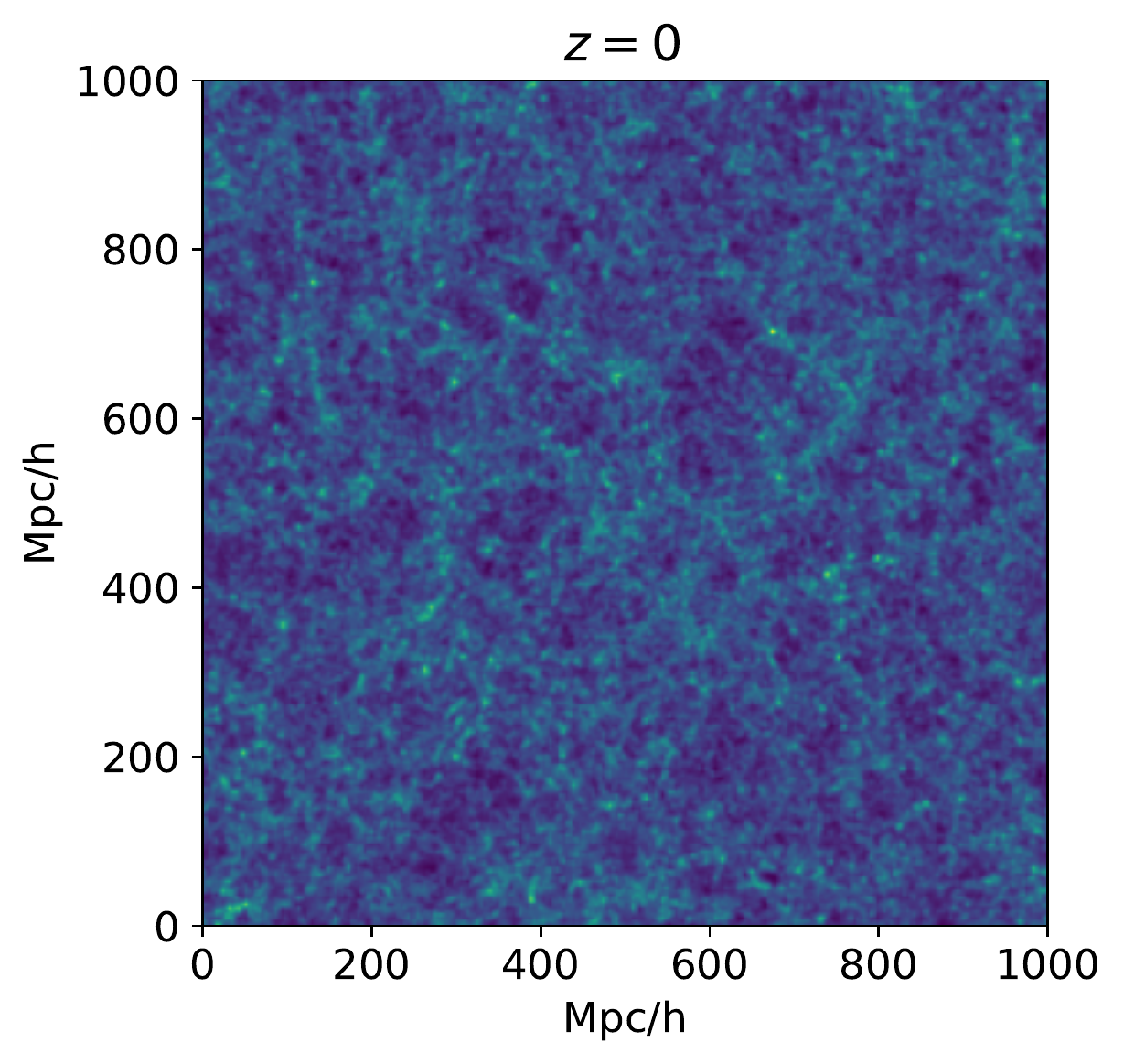}
    \includegraphics[scale=.61]{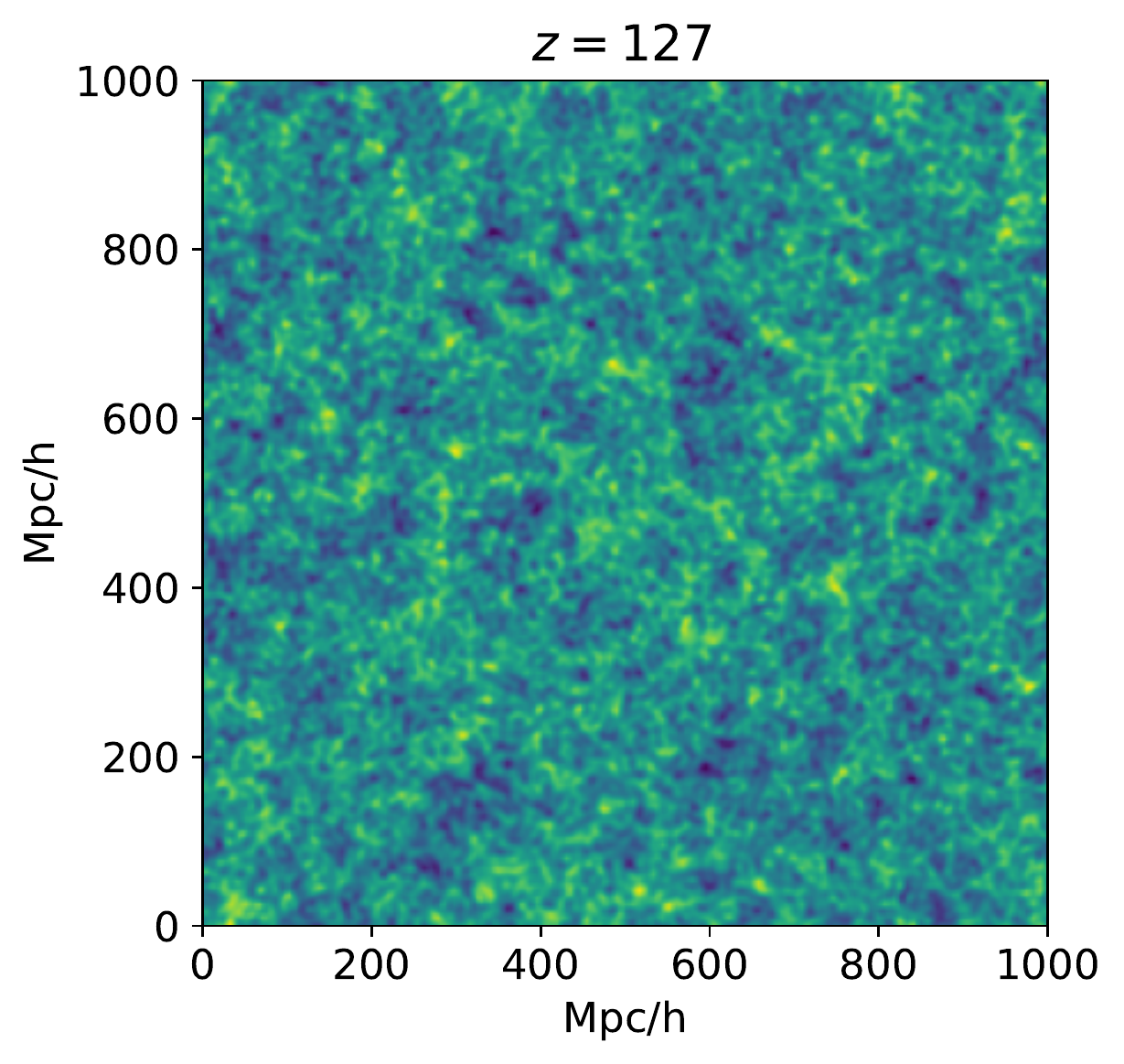}
    \includegraphics[scale=.61]{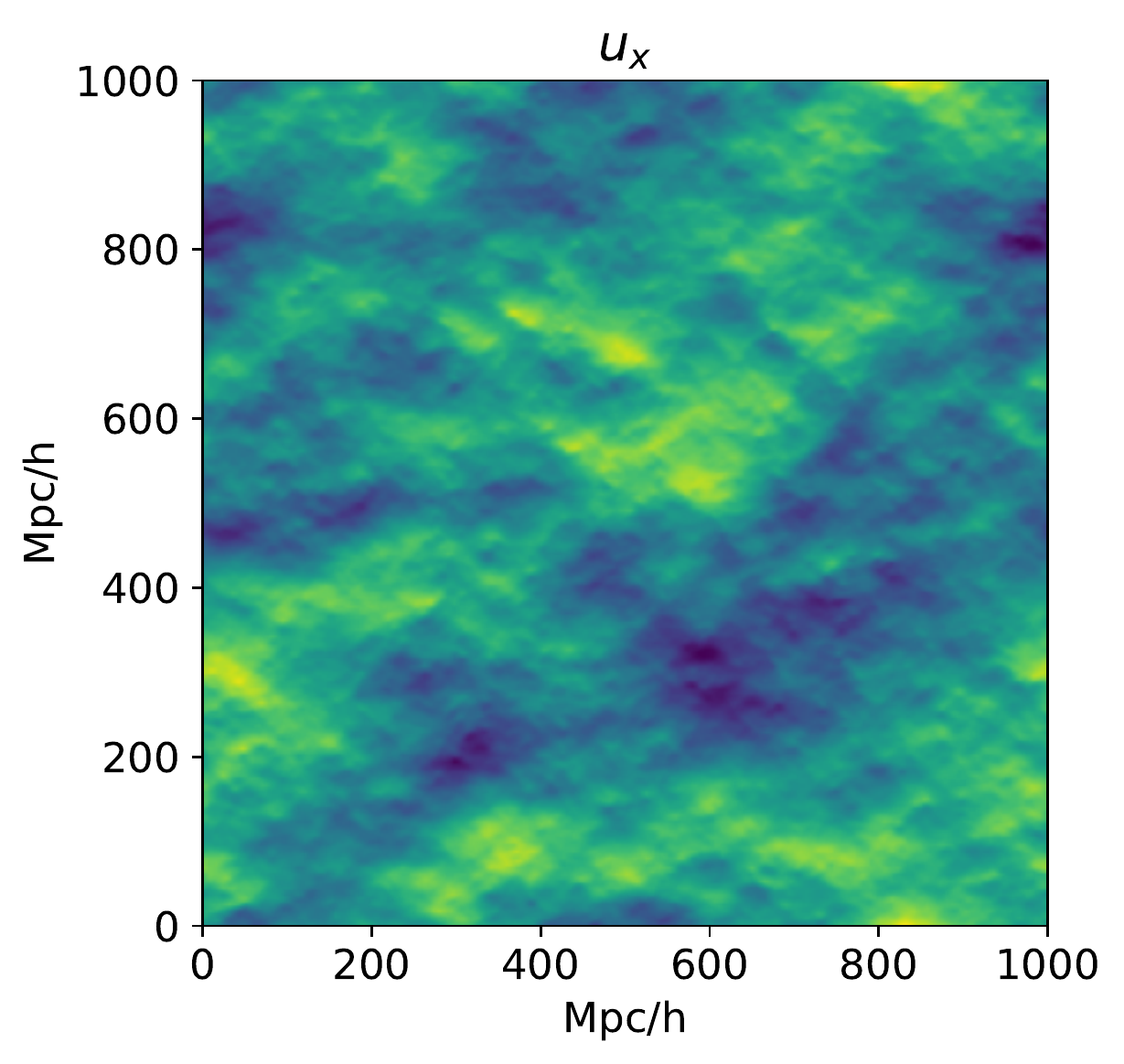}
    \includegraphics[scale=.61]{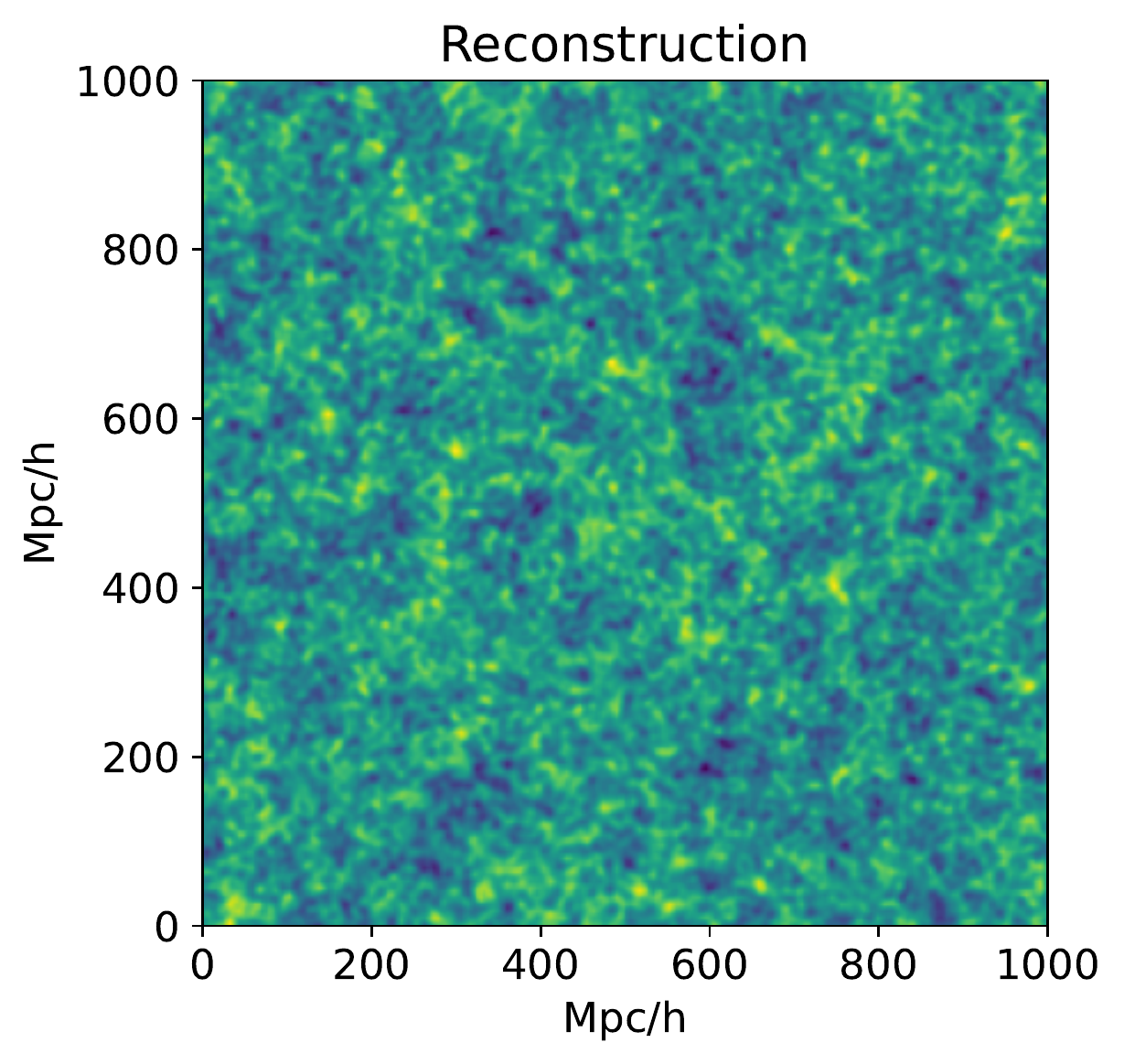}
    \caption{Top: a validation pair of final ($z=0$) and initial ($z=127$) density field. Bottom left: one of the velocity fields generated on the fly. Bottom right: the corresponding reconstruction produced by our network. These are averaged along one entire spatial dimension ($1000$ Mpc/h) for illustrative purposes, whereas the actual fields are three-dimensional.}
    \label{fig:2D_real}
\end{figure}

\begin{figure}
    \centering
    \includegraphics[scale=.51]{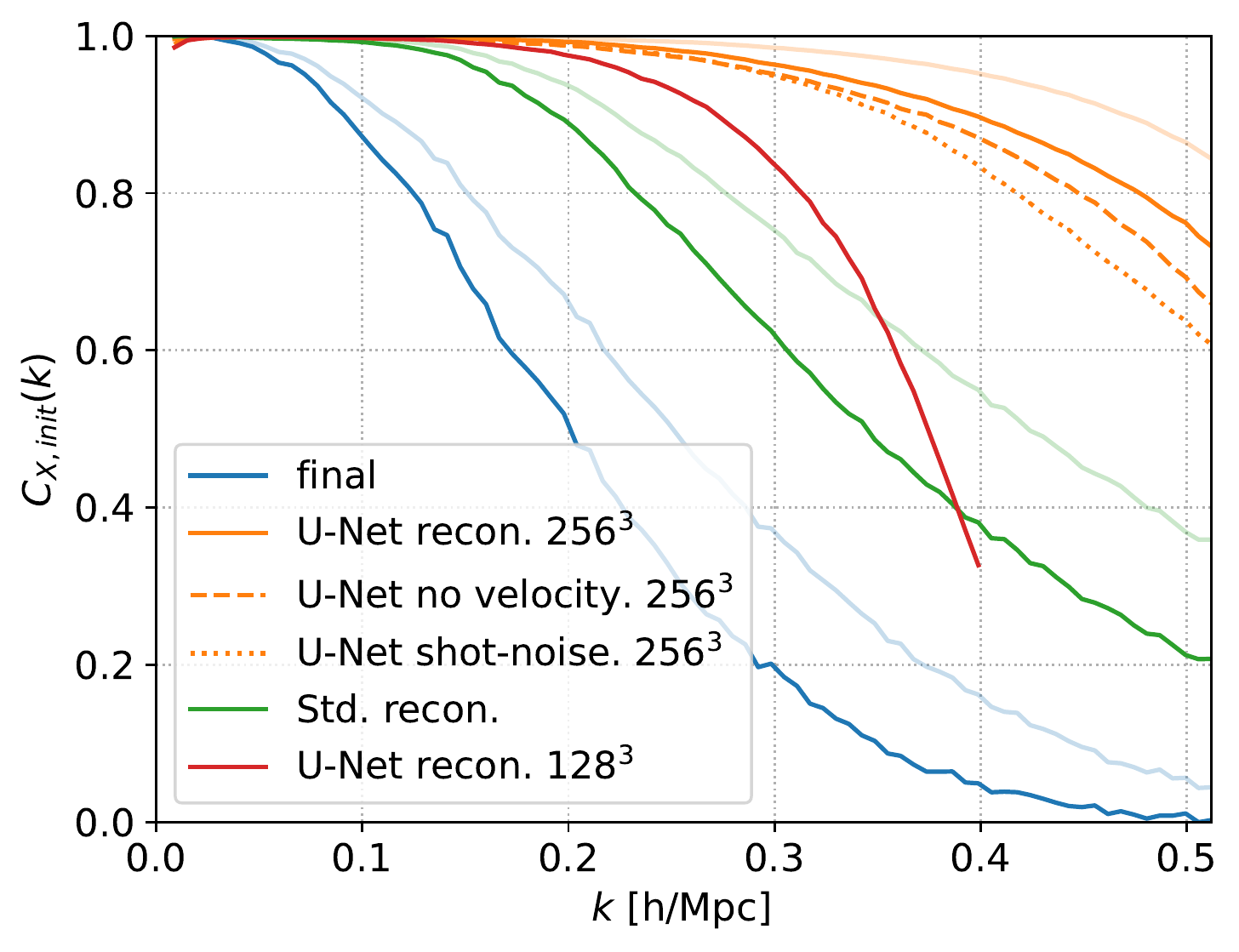}
    \includegraphics[scale=.51]{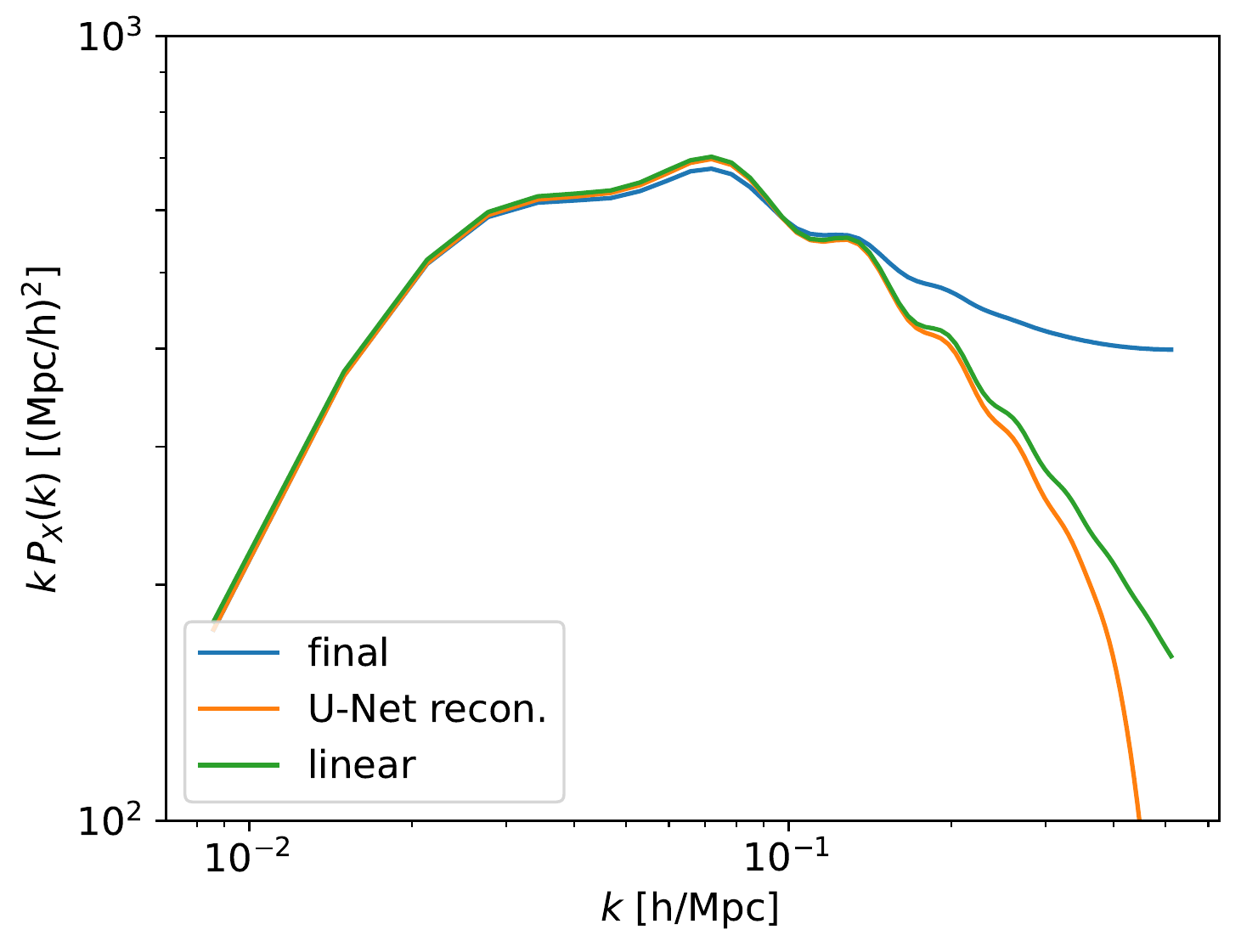}
    \caption{Left: Cross correlations of various density fields. Blue shows the cross-correlation of the final ($z=0$) and initial ($z=127$) density field. Solid orange shows the cross-correlation of the reconstructed and initial density field. Green shows the cross-correlation of the density field reconstructed using standard reconstruction with a smoothing scale of $\sigma = 10$ h/Mpc. Red shows the cross-correlation that can be achieved with a similar network and density fields of only $128^3$ cells. Dashed orange shows the performance of a network that only takes the density field as an input, without velocity fields. Dotten orange shows the performance of our network when the final density field contains increased shot noise (generated with 8 times fewer dark matter particles). Transparent lines show the same results but for final density fields at $z=0.5$ and can therefore be compared to Ref.~\cite{Shallue:2022mhf}. Right: mean power spectrum over 12500 simulations before and after reconstruction, compared to the linear power spectrum (note that we multiplied by an additional factor of $k$ to emphasize the BAO peaks).}
    \label{fig:crosscorr_real}
\end{figure}

We assess the quality of the reconstruction by computing the cross-correlation between two fields:
\begin{align}
    C_{X,Y}(\bk{}) = \frac{\langle \delta_{\rm X}(\bk{}) \delta_{\rm Y}^{*}(\bk{}) \rangle}{\sqrt{P_{\rm X}(\bk{})P_{\rm Y}(\bk{})}},
\end{align}
where $P_X(\bk{})$ is the power spectrum of the field $X$ and $X,Y$ can be either the reconstructed, initial ($z=127$) or final ($z=0$) density fields. A perfect reconstruction then corresponds to $C_{\rm recon, init}(\bk{}) = 1$. In Figure \ref{fig:2D_real} we show an example of the network input and output and its corresponding cross-correlations are shown in Figure \ref{fig:crosscorr_real}a, demonstrating the ability of our network to reconstruct the density field with a cross-correlation of $\sim 75\%$ up to the smallest scales of interest in this analysis. We also show the results if we use lower resolution density fields of $128^3$ cells, demonstrating that the main limitation to further improve the reconstruction is the resolution of the density fields, as the reconstruction benefits from having access to more small-scale modes that are coupled to larger ones that are to be reconstructed. We are currently limited to $256^3$ density fields due to memory limitations on our GPUs.\footnote{The approach taken in Ref.~\cite{Shallue:2022mhf} circumvents this issue by first performing standard reconstruction on a box of higher resolution and then reconstructing subboxes of smaller size. Since their analysis is performed at $z=0.5$, in Figure \ref{fig:crosscorr_real} we also plot our network's performance at that redshift for comparison. Note that these results have been obtained using different N-body simulations, fiducial cosmology, and resolution and should therefore only be compared to ours qualitatively.} Furthermore, we show the results for a similar network that does not use the velocity field information, demonstrating that including velocity is advantageous. Finally, we show the performance of our network when the $z=0$ density field contains an increased amount of shot-noise (we have generated the same density field using an 8 times smaller subsample of the dark matter particles), which degrades the reconstruction as expected \cite{Chen:2023uup}. Compared to the state-of-the-art iterative reconstruction algorithm of Ref.~\cite{Schmittfull:2017uhh}, our network achieves a similar accuracy (cross-correlation) of $\sim 90 \%$ up to scales $k\leq 0.4$ h/Mpc, while only having access to modes up to the Nyquist frequency of the $256^3$ grid, $k_{\rm Nyq} \approx 0.8$ h/Mpc. To further assess the ability of our network to `linearize' the density fields, we can consider the output power spectra. In Figure \ref{fig:crosscorr_real}b we show the mean power spectra over 12500 simulation boxes before and after reconstruction. We see that the reconstructed power spectrum more closely follows the linear power spectrum, especially recovering the BAO peaks. It is worth emphasizing that once trained, our network is able to perform reconstruction in under a second of time per field. Furthermore, our method works directly at the level of the density field, without having to perform computations on large object catalogs. This allows us to quickly reconstruct the large dataset necessary for our subsequent analysis of the information content of the pre and post-reconstructed power spectrum and bispectrum.

\section{Information content of the reconstructed density field}
\label{sec:INFO}
In order to determine the merit of our reconstruction for cosmological parameter inference, we perform a Fisher analysis of the information content of the power spectrum and bispectrum using the \texttt{QUIJOTE} simulation suite. These simulations allow us to work with the full non-linear power spectrum and bispectrum within the non-linear regime, giving us a reliable estimate of the improvements of parameter constraints, without the need for complicated models such as perturbation theory or effective field theory.

\subsection{Power spectrum and bispectrum estimators}
We measure the binned power spectrum from density fields $\delta(\bx{})$ according to:
\begin{align}
    \hat{P}(k_i) = \frac{1}{N_i} \sum_{\bq{} \in k_i} \delta(\bq{}) \delta^*(\bq{}),
\end{align}
where the sum runs over all momenta $\bq{}$ that are within the shell $[k_i - \Delta k /2,k_i + \Delta k /2)$ and $N_i$ is a normalization factor that counts the number of modes that fall in the bin. Similarly, for the bispectrum we compute
\begin{align}
    \hat{B}(k_1,k_2,k_3) = \frac{1}{N_{123}} \sum_{\bq{1} \in k_1}\sum_{\bq{2} \in k_2}\sum_{\bq{3} \in k_3}(2\pi)^3 \delta^{(3)}_D(\bq{1}+ \bq{2}+ \bq{3})\delta(\bq{1})\delta(\bq{2})\delta(\bq{3}),
\end{align}
where $N_{123}$ counts the number of triangle configurations within the bin. The bispectrum can be efficiently computed using fast Fourier transforms (FFTs). For the power spectrum, we use bins of width $\Delta k=k_F$ starting at the fundamental mode $k_{\rm min} = k_F$. The bispectrum is binned with width $\Delta k = 3k_F$, the first bin starting at $k_{\rm min} = 1.5k_F$. The smallest scale included in our analysis is $k_{\rm max} = 82.5k_F \approx 0.52$ h/Mpc resulting in $2276$ triangle bins. We apply the usual compensation for the mass assignment (PCS) window function in Fourier space before measuring our statistics \cite{Jing:2004fq}. Measuring the power spectrum and bispectrum of the pre and post-reconstructed density fields, we obtain our data products $\boldsymbol{D}=\{P_{\rm pre}, B_{\rm pre}, P_{\rm post}, B_{\rm post}\}$.\footnote{We have used our own code that was cross-validated with other bispectrum codes. Our code is publicly available at \url{https://github.com/tsfloss/DensityFieldTools}}\\

\subsection{Covariance}
\begin{figure}
    \centering
    \includegraphics[scale=.75]{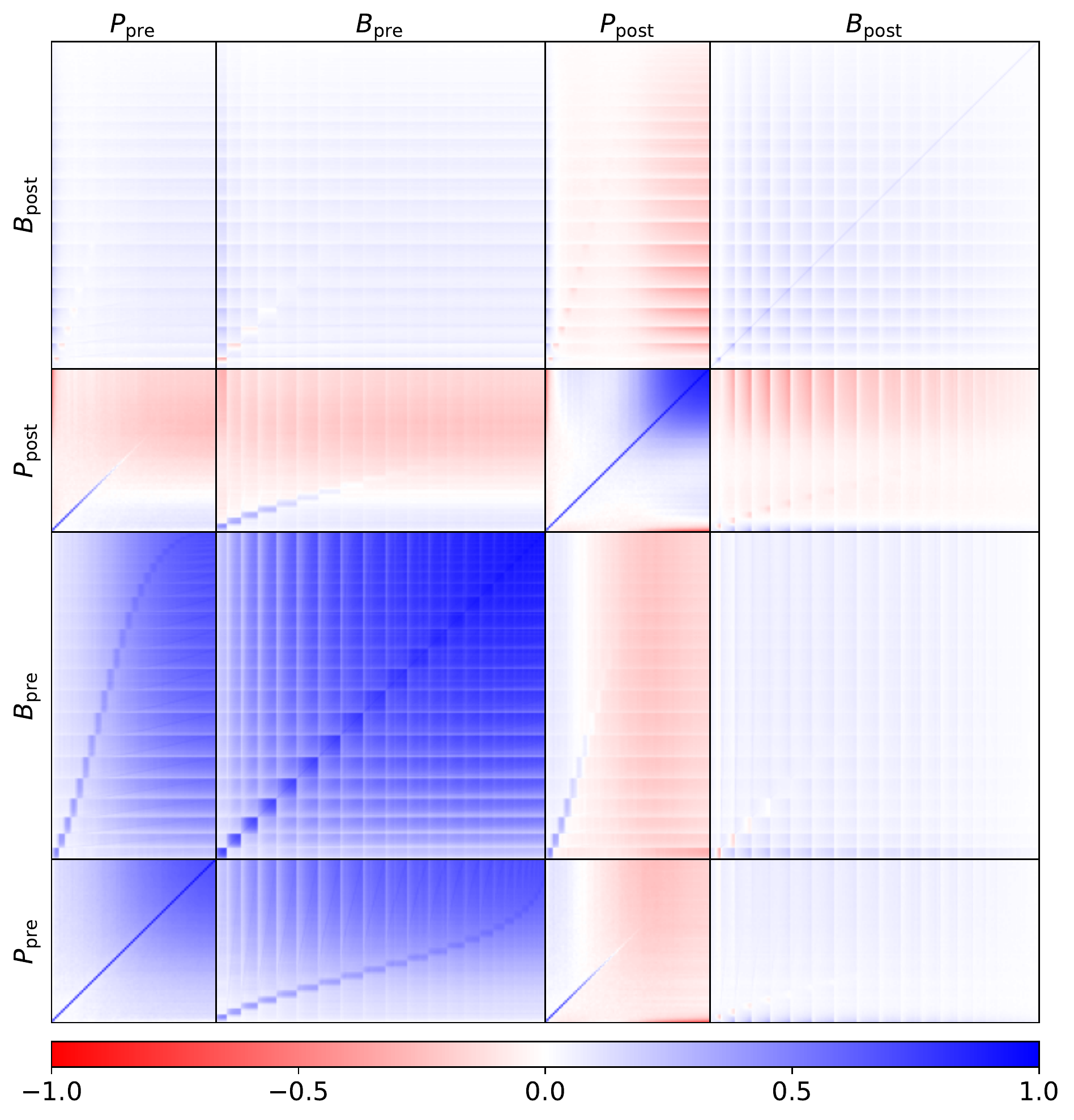}
    \caption{Correlation matrix $r_{ij}$ for the pre and post-reconstructed power spectrum and bispectrum. Triangle configurations for the bispectrum are ordered with increasing smallest momentum, visually creating blocks of triangles sharing the same shortest side in $k$-space.}
    \label{fig:cov}
\end{figure}

An important quantity for determining the information content is the covariance matrix, which encodes correlations between data points. The covariance matrix for a data vector $\boldsymbol{D}_n$ (in our case the power spectrum and bispectrum measurements) measured from simulation $n$, is given by:
\begin{align}
    C_{ij} = \frac{1}{N-1}\sum_n^{N} \left(\boldsymbol{D}_n - \bar{\boldsymbol{D}}  \right)_i\left(\boldsymbol{D}_n - \bar{\boldsymbol{D}}  \right)_j,
\end{align}
where $N$ is the number of measurements/simulations and $\bar{\boldsymbol{D}}$ is the mean of the data over all $N$ simulations. We obtain the reconstructed covariance matrix by measuring the power spectrum and bispectrum of $12500$ reconstructed \texttt{QUIJOTE} fiducial cosmology density fields at $z=0$, which we demonstrate to be sufficient in Appendix \ref{app:convergence}. In order to visualize the correlation between different power spectrum and bispectrum bins, we use the correlation matrix:
\begin{align}
    r_{ij} = \frac{C_{ij}}{\sqrt{C_{ii}C_{jj}}}.
\end{align}
In Figure \ref{fig:cov} we show the correlation matrices of the pre-and post-reconstructed power spectrum and bispectrum. In this figure, the triangle bins have been ordered with increasing smallest momenta (largest scale), which emphasizes the fact that before reconstruction, triangles sharing the same shortest side are severely correlated, especially when very squeezed \cite{Biagetti:2021tua}. After reconstruction, modes have become significantly less correlated, indicating that additional information from higher-order correlation functions has been brought into the reconstructed power spectrum and bispectrum. Since non-Gaussian covariance is responsible for the saturation of parameter constraints in the non-linear regime \cite{Biagetti:2021tua,Coulton:2022qbc,Floss:2022wkq}, we expect that the reduction of covariance due to our reconstruction method will translate into improved parameter constraints.

\subsection{Fisher Analysis}
We determine the constraining power of the reconstructed power spectrum and bispectrum by performing a Fisher analysis. Under the assumption of a Gaussian likelihood, which we expect to be accurate enough for our purpose \cite{Scoccimarro:2000sn}, the Fisher information matrix for parameters $\theta_a$ and data $\bar{\boldsymbol{D}}$ (e.g. power spectrum and/or bispectrum) is given by:
\begin{align}
    \label{eq:fish}
    F_{ab} = \frac{\partial \bar{\boldsymbol{D}}}{\partial \theta_a}\cdot \left(C^{-1}\right) \cdot \frac{\partial \bar{\boldsymbol{D}}}{\partial \theta_b},
\end{align}
where the dot product runs over the data points. The inverse covariance matrix or precision matrix $C^{-1}$, is obtained by numerically inverting the covariance matrix of the previous section. The inversion of covariance matrices is biased by the limited amount of realizations available to construct the covariance matrix. We unbias our precision matrix by including the Hartlap factor \cite{Hartlap:2006kj}. In order to obtain the derivatives with respect to the cosmological parameters that we would like to constrain, we make use of the \texttt{QUIJOTE} simulations that vary $\{ f^{\rm local}_{\rm NL},f^{\rm equil}_{\rm NL},f^{\rm orth}_{\rm NL},h,n_s,\Omega_m,\Omega_b,\sigma_8\}$ by a fixed amount around their fiducial values, i.e. $\theta^{\rm fid}_a \pm \delta \theta_a$  as summarized in Table \ref{tab:quijote} \cite{Villaescusa-Navarro:2019bje,Coulton:2022qbc}. We reconstruct all these simulations (500 per varied parameter) and measure their power spectra and bispectra. We then compute the parameter derivatives by central differencing:
\begin{align}
\label{eq:Dderiv}
    \frac{\partial \bar{\boldsymbol{D}}}{\partial \theta_a} = \frac{\bar{\boldsymbol{D}}_{\theta^{\rm fid}_a + \delta \theta_a}-\bar{\boldsymbol{D}}_{\theta^{\rm fid}_a - \delta \theta_a}}{2\delta \theta_a}.
\end{align}
Here $\bar{\boldsymbol{D}}_{\theta^{\rm fid}_a \pm \delta \theta_a}$ presents the mean measured data from the simulations run with cosmological parameter $\theta_a = \theta^{\rm fid}_a \pm \delta \theta_a$. The Fisher matrix allows us to compute the marginalized error on model parameters: 
\begin{table}[]
\centering
\begin{tabular}{|l|l|l|l|l|l|l|l|l|}
\hline
                    & $f^{\rm loc}_{\rm NL}$ & $f^{\rm equil}_{\rm NL}$ & $f^{\rm orth}_{\rm NL}$& $h$ & $n_s$ & $\Omega_m$ & $\Omega_b$ & $\sigma_8$  \\ \hline
$\theta^{\rm fid}$            & 0 & 0 & 0 & $0.6711$ & $0.9624$ & $0.3175$ & $0.049$ & $0.834$ \\ \hline
$\delta \theta$     & $\pm 100$ & $\pm 100$ & $\pm 100$ & $\pm 0.02$ & $\pm 0.02$ & $\pm 0.01 $ & $\pm 0.002$ & $\pm 0.015$ \\ \hline
\end{tabular}
\caption{Cosmological parameters of the \texttt{QUIJOTE} simulations used in this paper. The top row gives the fiducial cosmology used for training the neural network and computing the covariance matrix. The bottom row gives the variation of the parameters in simulations used to compute the derivatives in equation~\eqref{eq:Dderiv}}
\label{tab:quijote}
\end{table}
\begin{align}
    \sigma_a = \sqrt{(F^{-1})_{aa}},
\end{align}
as well as their correlation coefficient, quantifying the amount of degeneracy between parameters:
\begin{align}
    \rho_{ab} = \frac{(F^{-1})_{ab}}{\sqrt{\sigma^2_a \sigma^2_b}}.
\end{align}
Furthermore, the Fisher matrix can be used to construct an unbiased, minimal-variance marginalized estimator that estimates the parameter $\theta_a$ from the observed data $\boldsymbol{D}^{\rm obs}$ given a fiducial model $\bar{\boldsymbol{D}}^{\rm fid}$:
\begin{align}
\label{eq:estimator}
    \hat{\theta}_a - \theta^{\rm fid}_{a}= \sum_{b} (F^{-1})_{ab} ~ \frac{\partial \bar{\boldsymbol{D}}}{\partial \theta_b} \cdot \left( C^{-1} \right)\cdot \left(\boldsymbol{D}^{\rm obs} - \bar{\boldsymbol{D}}^{\rm fid}\right).
\end{align}
The expected error of the estimate is equal to the Fisher error $\langle \hat{\theta}^2_a \rangle = \sigma^2_a = (F^{-1})_{aa}$. This estimator marginalizes over all parameters to make sure that parameter discrepancies between our fiducial model and the real cosmology do not translate into potentially large biases of the estimated parameter. Since our network is trained only on the fiducial cosmology simulations, reconstructions of simulations with different parameters are biased. Since we use these biased reconstructions to compute the derivatives in equation \eqref{eq:Dderiv}, this bias is thus included in the marginalization process. We therefore expect estimates of primordial non-Gaussianity to be unbiased and minimal-variance, even when the cosmology of the measured data does not match the fiducial cosmology. \\

\subsection{Results}
\begin{figure}
    \centering
    \includegraphics[scale=.70]{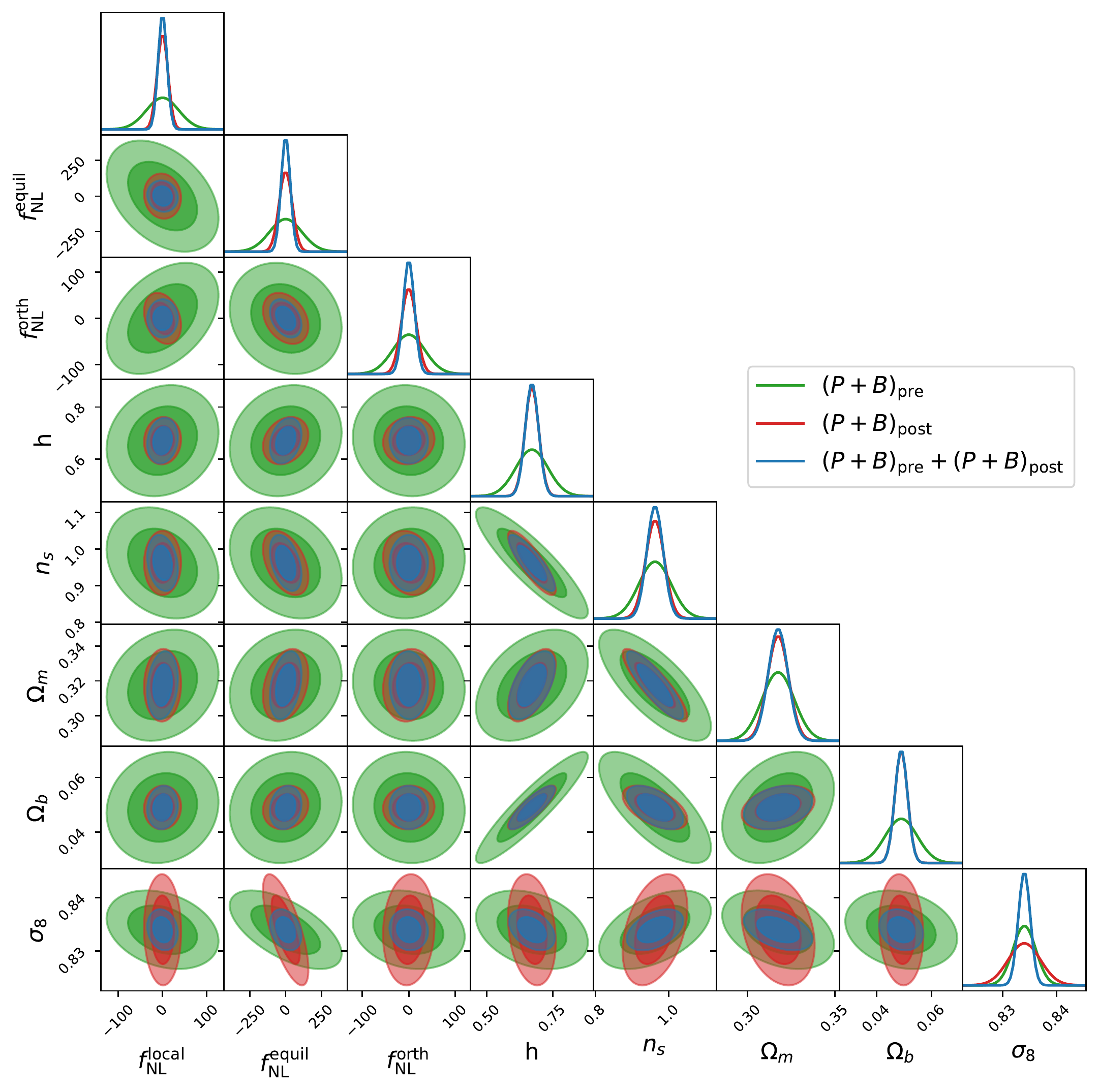}
    \caption{1 and 2-$\sigma$ marginalized constraints when jointly fitting for all eight parameters using the power spectrum and bispectrum \textcolor{tabgreen}{pre}, \textcolor{tabred}{post} and \textcolor{tabblue}{pre+post} reconstruction.}
    \label{fig:confidenceplot}
\end{figure}

\begin{figure}
    \centering\includegraphics[scale=.6]{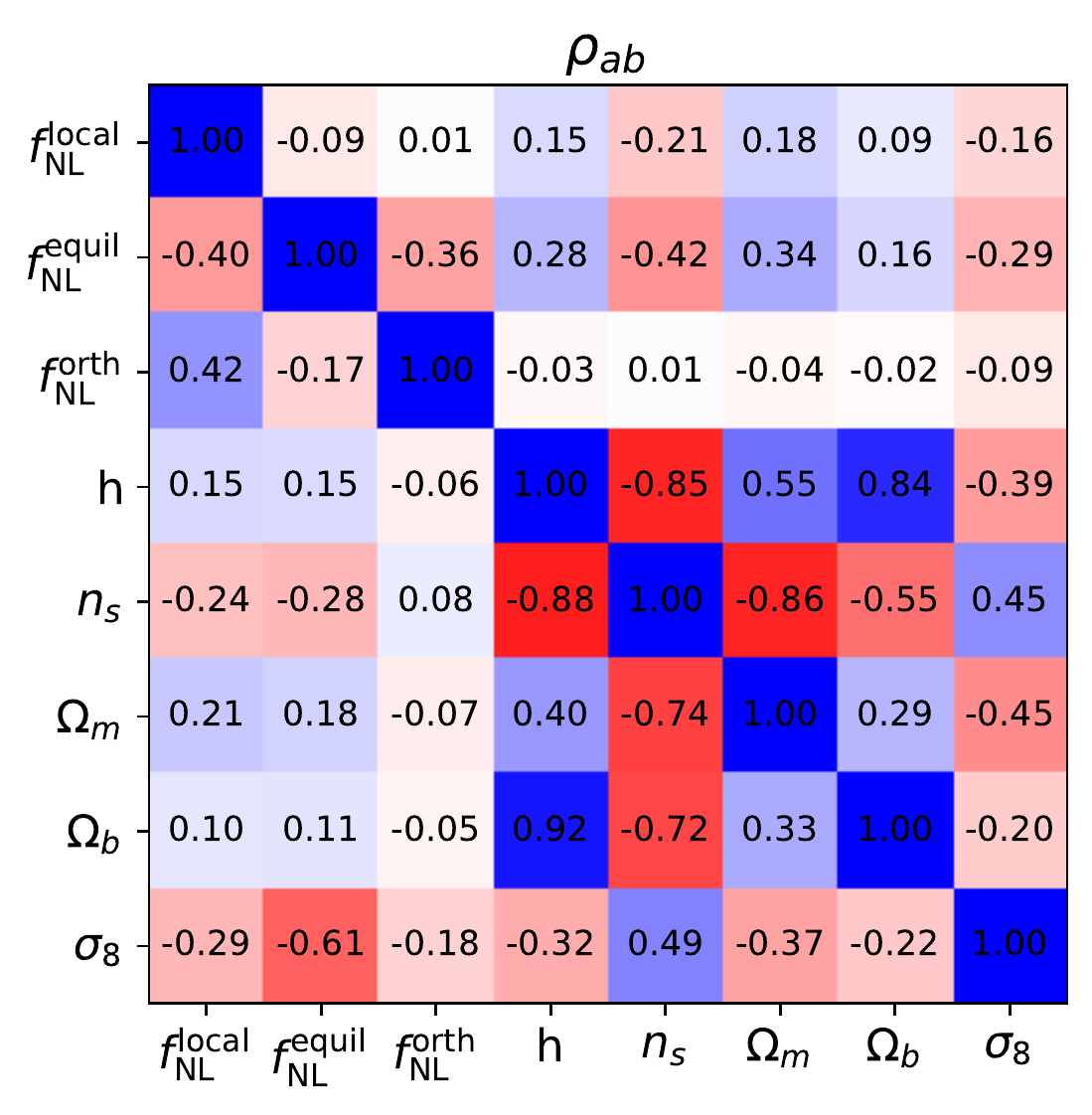}
    \caption{Correlation coefficients $\rho_{ij} \in [-1,1]$ between the different cosmological parameters when using the data products $(P+B)_{\rm pre}$ (lower triangular) and $(P+B)_{\rm pre} + (P+B)_{\rm post}$ (upper triangular).}
    \label{fig:rho}
\end{figure}

From the calculation of the Fisher information matrix of the pre and post-reconstructed power spectrum and bispectrum, Figure \ref{fig:confidenceplot} shows the resulting confidence intervals and marginalized errors of all 8 parameters when fitting for all of them simultaneously. The results are based on using both power spectrum and bispectrum data (see Appendix~\ref{app:PandB} for the results when using the power spectrum and bispectrum separately). The results before reconstruction match those presented in Ref.~\cite{Coulton:2022qbc}. Most notably, after reconstruction there is a significant improvement in the error of all parameters when combining the pre- and post-reconstructed statistics as summarized in Table \ref{tab:improvement}. To understand the improvement of these constraints, Figure \ref{fig:rho} shows the correlation coefficient between the parameter constraints, revealing a drastic reduction in degeneracy model parameters. In Table \ref{tab:fnl}, we present the (un-)marginalized constraints on $f_{\rm NL}$. We conclude that our reconstruction method improves marginalized constraints on $f_{\rm NL}$ by a factor of $3.65$, $3.54$, and $2.90$ for local, equilateral, and orthogonal respectively. The improved constraints are attributed to the reduced covariance and parameter degeneracy. Our results show that in order to get an accurate estimate of the improvement on $f_{\rm NL}$ it is important to include both the power spectrum and bispectrum and to marginalize over other cosmological parameters. Naively, using only post-reconstructed bispectrum measurements ($B_{\rm post}$) would lead to an improvement of marginalized constraints by a factor of $5.55$ (local), $2.58$ (equilateral) and $3.12$ (orthogonal). Similarly, using only the tree-level primordial bispectrum contribution in equation \eqref{eq:treeprimB}, as was done in Ref.~\cite{Shirasaki:2020vkk}, suggests improvements on marginalized constraints by a factor $11.69$ (local), $4.76$ (equilateral) and $2.14$ (orthogonal), demonstrating the necessity of including the full non-linear power spectrum and bispectrum in order to arrive at realistic estimates. Finally, estimating $f_{\rm NL}$ from the (reconstructed) \texttt{QUIJOTE} simulations using the estimator in equation \eqref{eq:estimator} with the different products, we confirm that the estimator is unbiased and minimum-variance.

\begin{table}[]
\centering 
\begin{tabular}{llllllll}
$f_{\rm NL}^{\rm local}$ & $f_{\rm NL}^{\rm equil}$ & $f_{\rm NL}^{\rm orth}$ & $h$ & $n_s$ & $\Omega_m$ & $\Omega_b$ & $\sigma_8$ \\\hline
3.65 & 3.54 & 2.90 & 2.43 & 1.98 & 1.64 & 2.56 & 1.91 \\ \hline
\end{tabular}
\caption{Improvement factor of parameter constraints when using the full data product $(P+B)_{\rm pre}+(P+B)_{\rm post}$ as compared to just $(P+B)_{\rm pre}$}
\label{tab:improvement}
\end{table}

 \begin{table}[]
 \centering
\begin{tabular}{|p{44mm}|p{22mm}|p{22mm}|p{22mm}|}
  \hline
  & $f_{\rm NL}^{\rm local}$ & $f_{\rm NL}^{\rm equil}$ & $f_{\rm NL}^{\rm orth}$ \\ \hline
  \textcolor{tabgreen}{$P_{\rm pre}^*$} \newline \textcolor{tabred}{$P_{\rm post}$} \newline \textcolor{tabblue}{$P_{\rm pre} + P_{\rm post}$} & \textcolor{tabgreen}{2884 (31.31)} \newline \textcolor{tabred}{1890 (69.11)} \newline \textcolor{tabblue}{135.6 (29.06)} & \textcolor{tabgreen}{7888 (79.41)} \newline \textcolor{tabred}{3500 (165.5)} \newline \textcolor{tabblue}{481.2 (76.94)} & \textcolor{tabgreen}{3478 (187.2)} \newline \textcolor{tabred}{572.8 (116.4)} \newline \textcolor{tabblue}{175.4 (77.70)} \\ \hline
    \textcolor{tabgreen}{$B_{\rm pre}$} \newline \textcolor{tabred}{$B_{\rm post}$} \newline \textcolor{tabblue}{$B_{\rm pre} + B_{\rm post}$} & \textcolor{tabgreen}{101.8 (42.65)} \newline\textcolor{tabred}{18.35 (16.86)} \newline \textcolor{tabblue}{16.94 (15.61)}& \textcolor{tabgreen}{187.7 (123.8)} \newline \textcolor{tabred}{72.63 (34.25)} \newline \textcolor{tabblue}{60.65 (33.05)}& \textcolor{tabgreen}{83.39 (61.39)} \newline \textcolor{tabred}{26.76 (16.85)} \newline \textcolor{tabblue}{22.03 (15.51)}\\ \hline
      \textcolor{tabgreen}{$(P+B)_{\rm pre}$} \newline \textcolor{tabred}{$(P+B)_{\rm post}$} \newline \textcolor{tabblue}{$(P+B)_{\rm pre}+(P+B)_{\rm post}$} & \textcolor{tabgreen}{51.64 (24.45)} \newline \textcolor{tabred}{17.10 (15.61)} \newline \textcolor{tabblue}{14.14 (12.77)} &\textcolor{tabgreen}{159.3 (59.77)} \newline \textcolor{tabred}{64.73 (32.58)} \newline \textcolor{tabblue}{45.01 (30.21)}& \textcolor{tabgreen}{49.08 (42.56)} \newline \textcolor{tabred}{22.58 (15.49)} \newline \textcolor{tabblue}{16.93 (13.90)}\\ \hline
\end{tabular}
\caption{Marginalized (Unmarginalized) errors on $f_{\rm NL}$ \textcolor{tabgreen}{pre}, \textcolor{tabred}{post} and \textcolor{tabblue}{pre+post} reconstruction using different data products for a dark matter survey with volume $V = 1$ (Gpc/h)$^3$ at redshift $z=0$ and $k_{\rm max} \approx 0.52$ h/Mpc. $^*$ Note that the parameter constraints from the pre-reconstructed power spectrum alone are likely too optimistic since the Fisher derivatives have not converged, as discussed in Appendix \ref{app:convergence} and Ref.~\cite{Coulton:2022qbc}.}
\label{tab:fnl}
 \end{table}

\section{Conclusions}\label{sec:conclusion}
In this work, we have studied the use of neural networks in reconstructing the initial conditions from the late-time density field, with the main goal of improving future constraints on primordial non-Gaussianity. We have demonstrated that a U-Net architecture is able to reconstruct the initial conditions with an accuracy (cross-correlation) of $\sim 90 \%$ up to scales $k \leq 0.4$ h/Mpc directly from the redshift $z=0$ density field at a fraction of the computational time required by competitive methods such as that of Ref.~\cite{Schmittfull:2017uhh}. Furthermore, we have analyzed the information content of the dark matter density field after reconstruction by reconstructing a large part of the \texttt{QUIJOTE} simulation suite. From a Fisher analysis, we find that marginalized constraints on primordial non-Gaussianity from a joint analysis of power spectrum and bispectrum, are improved by a factor of $3.65$, $3.54$, $2.90$ for local, equilateral, and orthogonal non-Gaussianity respectively. Given the strong saturation of these constraints in the non-linear regime due to non-Gaussian covariance \cite{Coulton:2022qbc}, these constitute valuable improvements that cannot be realized by just including $\sim 10$ times more modes, as a naive mode-counting argument suggests. Hence, our work marks important progress towards a more optimal estimation of primordial non-Gaussianity.\\

It is important to note that we have not performed any extensive optimization of our network architecture or its hyperparameters, thus it is conceivable that the reconstruction performance can be further improved. Furthermore, our reconstruction is so far limited by the amount of memory available on GPUs.\\
\\
This work can naturally be extended in a number of ways that we intend to explore in a future work:

\begin{enumerate}
    \item In this work we have limited ourselves to the dark matter density field. In reality, we only have access to a biased tracer thereof, such as galaxies or atomic/molecular spectral lines (e.g. 21-cm). At low redshifts, these trace dark matter halos. To further assess the realism of this method we need to apply our methods to the dark matter halo field and eventually a real tracer.
    \item In a real survey, we cannot observe the real position along the line-of-sight direction. Instead, we determine their position in redshift space, which is uncertain due to the unknown peculiar velocity of the objects. These \emph{redshift space distortions} are an additional source of non-linearity. Again, instead, we should provide our network with late-time density fields in redshift space as input and train this to reconstruct the real space initial conditions.
\end{enumerate}

Future constraints on primordial non-Gaussianity coming from large-scale structure will rely on our ability to model the non-linear nature of the tracer field. Traditionally, the emphasis in this field has been on modeling the statistical properties of the tracer field and trying to push these further into the non-linear regime in order to access more modes that can be compared to observation. Only more recently it has been shown that this approach shows diminishing returns as non-linear modes become increasingly correlated, reducing their information content. This is essentially a manifestation of the non-linear evolution scrambling the information of the linear field into higher-order correlation functions and additional work needs to be done to recover the information contained in the initial conditions. Therefore, it is important to emphasize that the complications due to mode coupling encountered in this work essentially turn N-point correlation functions into highly degenerate sub-optimal statistics. Although correlation functions are the natural object to study from a theoretical perspective, especially when it comes to primordial non-Gaussianity, there might be more optimal ways to compress survey data into statistics (see e.g. \cite{Valogiannis:2021chp,Jung:2022rtn}). It would be interesting to study alternative statistics in the context of reconstruction as well.

\section*{Acknowledgements}
The authors would like to thank Matteo Biagetti, William Coulton, and Francisco Villaescusa-Navarro for useful discussions related to this work and for comments on the manuscript. We would also like to thank Francisco Villaescusa-Navarro, William Coulton, and the whole \textsc{QUIJOTE} team for making the simulation suite available \cite{Villaescusa-Navarro:2019bje}. We thank the Center for Information Technology of the University of Groningen for providing access to the Peregrine and Hábrók high-performance computing clusters. T.F. is supported by the Fundamentals of the Universe research program at the University of Groningen. P.D.M. acknowledges support from the Netherlands Organization for Scientific Research (NWO) VIDI grant (dossier 639.042.730).

\bibliographystyle{JHEP}
\bibliography{bibliography.bib}

\providecommand{\href}[2]{#2}\begingroup\raggedright\begin{thebibliography}{10}

\bibitem{Achucarro:2022qrl}
A.~Ach\'ucarro et~al., \emph{{Inflation: Theory and Observations}},
  \href{http://arxiv.org/abs/2203.08128}{{\tt 2203.08128}}.

\bibitem{Meerburg:2019qqi}
P.~D. Meerburg et~al., \emph{{Primordial Non-Gaussianity}},
  \href{http://arxiv.org/abs/1903.04409}{{\tt 1903.04409}}.

\bibitem{Arkani-Hamed:2015bza}
N.~Arkani-Hamed and J.~Maldacena, \emph{{Cosmological Collider Physics}},
  \href{http://arxiv.org/abs/1503.08043}{{\tt 1503.08043}}.

\bibitem{Lee:2016vti}
H.~Lee, D.~Baumann and G.~L. Pimentel, \emph{{Non-Gaussianity as a Particle
  Detector}}, \href{http://dx.doi.org/10.1007/JHEP12(2016)040}{\emph{JHEP} {\bf
  12} (2016) 040}, [\href{http://arxiv.org/abs/1607.03735}{{\tt 1607.03735}}].

\bibitem{Ferreira:1998kt}
P.~G. Ferreira, J.~Magueijo and K.~M. Gorski, \emph{{Evidence for
  nonGaussianity in the DMR four year sky maps}},
  \href{http://dx.doi.org/10.1086/311514}{\emph{Astrophys. J. Lett.} {\bf 503}
  (1998) L1--L4}, [\href{http://arxiv.org/abs/astro-ph/9803256}{{\tt
  astro-ph/9803256}}].

\bibitem{Komatsu:2001wu}
E.~Komatsu, B.~D. Wandelt, D.~N. Spergel, A.~J. Banday and K.~M. Gorski,
  \emph{{Measurement of the cosmic microwave background bispectrum on the COBE
  DMR sky maps}}, \href{http://dx.doi.org/10.1086/337963}{\emph{Astrophys. J.}
  {\bf 566} (2002) 19--29}, [\href{http://arxiv.org/abs/astro-ph/0107605}{{\tt
  astro-ph/0107605}}].

\bibitem{WMAP:2003xez}
{\scshape WMAP} collaboration, E.~Komatsu et~al., \emph{{First year Wilkinson
  Microwave Anisotropy Probe (WMAP) observations: tests of gaussianity}},
  \href{http://dx.doi.org/10.1086/377220}{\emph{Astrophys. J. Suppl.} {\bf 148}
  (2003) 119--134}, [\href{http://arxiv.org/abs/astro-ph/0302223}{{\tt
  astro-ph/0302223}}].

\bibitem{Creminelli:2005hu}
P.~Creminelli, A.~Nicolis, L.~Senatore, M.~Tegmark and M.~Zaldarriaga,
  \emph{{Limits on non-gaussianities from wmap data}},
  \href{http://dx.doi.org/10.1088/1475-7516/2006/05/004}{\emph{JCAP} {\bf 05}
  (2006) 004}, [\href{http://arxiv.org/abs/astro-ph/0509029}{{\tt
  astro-ph/0509029}}].

\bibitem{Planck:2015zfm}
{\scshape Planck} collaboration, P.~A.~R. Ade et~al., \emph{{Planck 2015
  results. XVII. Constraints on primordial non-Gaussianity}},
  \href{http://dx.doi.org/10.1051/0004-6361/201525836}{\emph{Astron.
  Astrophys.} {\bf 594} (2016) A17},
  [\href{http://arxiv.org/abs/1502.01592}{{\tt 1502.01592}}].

\bibitem{Planck:2019kim}
{\scshape Planck} collaboration, Y.~Akrami et~al., \emph{{Planck 2018 results.
  IX. Constraints on primordial non-Gaussianity}},
  \href{http://dx.doi.org/10.1051/0004-6361/201935891}{\emph{Astron.
  Astrophys.} {\bf 641} (2020) A9},
  [\href{http://arxiv.org/abs/1905.05697}{{\tt 1905.05697}}].

\bibitem{SimonsObservatory:2018koc}
{\scshape Simons Observatory} collaboration, P.~Ade et~al., \emph{{The Simons
  Observatory: Science goals and forecasts}},
  \href{http://dx.doi.org/10.1088/1475-7516/2019/02/056}{\emph{JCAP} {\bf 02}
  (2019) 056}, [\href{http://arxiv.org/abs/1808.07445}{{\tt 1808.07445}}].

\bibitem{CMB-S4:2016ple}
{\scshape CMB-S4} collaboration, K.~N. Abazajian et~al., \emph{{CMB-S4 Science
  Book, First Edition}},  \href{http://arxiv.org/abs/1610.02743}{{\tt
  1610.02743}}.

\bibitem{Kalaja:2020mkq}
A.~Kalaja, P.~D. Meerburg, G.~L. Pimentel and W.~R. Coulton, \emph{{Fundamental
  limits on constraining primordial non-Gaussianity}},
  \href{http://dx.doi.org/10.1088/1475-7516/2021/04/050}{\emph{JCAP} {\bf 04}
  (2021) 050}, [\href{http://arxiv.org/abs/2011.09461}{{\tt 2011.09461}}].

\bibitem{Munoz:2015eqa}
J.~B. Mu\~noz, Y.~Ali-Ha\"\i{}moud and M.~Kamionkowski, \emph{{Primordial
  non-gaussianity from the bispectrum of 21-cm fluctuations in the dark ages}},
  \href{http://dx.doi.org/10.1103/PhysRevD.92.083508}{\emph{Phys. Rev. D} {\bf
  92} (2015) 083508}, [\href{http://arxiv.org/abs/1506.04152}{{\tt
  1506.04152}}].

\bibitem{Meerburg:2016zdz}
P.~D. Meerburg, M.~M\"unchmeyer, J.~B. Mu\~noz and X.~Chen, \emph{{Prospects
  for Cosmological Collider Physics}},
  \href{http://dx.doi.org/10.1088/1475-7516/2017/03/050}{\emph{JCAP} {\bf 03}
  (2017) 050}, [\href{http://arxiv.org/abs/1610.06559}{{\tt 1610.06559}}].

\bibitem{Karagiannis:2020dpq}
D.~Karagiannis, J.~Fonseca, R.~Maartens and S.~Camera, \emph{{Probing
  primordial non-Gaussianity with the power spectrum and bispectrum of future
  21 cm intensity maps}},
  \href{http://dx.doi.org/10.1016/j.dark.2021.100821}{\emph{Phys. Dark Univ.}
  {\bf 32} (2021) 100821}, [\href{http://arxiv.org/abs/2010.07034}{{\tt
  2010.07034}}].

\bibitem{Floss:2022grj}
T.~Fl\"oss, T.~de~Wild, P.~D. Meerburg and L.~V.~E. Koopmans, \emph{{The Dark
  Ages' 21-cm trispectrum}},
  \href{http://dx.doi.org/10.1088/1475-7516/2022/06/020}{\emph{JCAP} {\bf 06}
  (2022) 020}, [\href{http://arxiv.org/abs/2201.08843}{{\tt 2201.08843}}].

\bibitem{Cabass:2022epm}
G.~Cabass, M.~M. Ivanov, O.~H.~E. Philcox, M.~Simonovic and M.~Zaldarriaga,
  \emph{{Constraining Single-Field Inflation with MegaMapper}},
  \href{http://arxiv.org/abs/2211.14899}{{\tt 2211.14899}}.

\bibitem{Pritchard:2011xb}
J.~R. Pritchard and A.~Loeb, \emph{{21-cm cosmology}},
  \href{http://dx.doi.org/10.1088/0034-4885/75/8/086901}{\emph{Rept. Prog.
  Phys.} {\bf 75} (2012) 086901}, [\href{http://arxiv.org/abs/1109.6012}{{\tt
  1109.6012}}].

\bibitem{Kovetz:2017agg}
E.~D. Kovetz et~al., \emph{{Line-Intensity Mapping: 2017 Status Report}},
  \href{http://arxiv.org/abs/1709.09066}{{\tt 1709.09066}}.

\bibitem{Bernardeau:2001qr}
F.~Bernardeau, S.~Colombi, E.~Gaztanaga and R.~Scoccimarro, \emph{{Large scale
  structure of the universe and cosmological perturbation theory}},
  \href{http://dx.doi.org/10.1016/S0370-1573(02)00135-7}{\emph{Phys. Rept.}
  {\bf 367} (2002) 1--248}, [\href{http://arxiv.org/abs/astro-ph/0112551}{{\tt
  astro-ph/0112551}}].

\bibitem{Taruya:2012ut}
A.~Taruya, F.~Bernardeau, T.~Nishimichi and S.~Codis, \emph{{RegPT: Direct and
  fast calculation of regularized cosmological power spectrum at two-loop
  order}}, \href{http://dx.doi.org/10.1103/PhysRevD.86.103528}{\emph{Phys. Rev.
  D} {\bf 86} (2012) 103528}, [\href{http://arxiv.org/abs/1208.1191}{{\tt
  1208.1191}}].

\bibitem{Carrasco:2012cv}
J.~J.~M. Carrasco, M.~P. Hertzberg and L.~Senatore, \emph{{The Effective Field
  Theory of Cosmological Large Scale Structures}},
  \href{http://dx.doi.org/10.1007/JHEP09(2012)082}{\emph{JHEP} {\bf 09} (2012)
  082}, [\href{http://arxiv.org/abs/1206.2926}{{\tt 1206.2926}}].

\bibitem{Slosar:2008hx}
A.~Slosar, C.~Hirata, U.~Seljak, S.~Ho and N.~Padmanabhan, \emph{{Constraints
  on local primordial non-Gaussianity from large scale structure}},
  \href{http://dx.doi.org/10.1088/1475-7516/2008/08/031}{\emph{JCAP} {\bf 08}
  (2008) 031}, [\href{http://arxiv.org/abs/0805.3580}{{\tt 0805.3580}}].

\bibitem{Ross:2012sx}
A.~J. Ross et~al., \emph{{The Clustering of Galaxies in SDSS-III DR9 Baryon
  Oscillation Spectroscopic Survey: Constraints on Primordial
  Non-Gaussianity}}, \href{http://dx.doi.org/10.1093/mnras/sts094}{\emph{Mon.
  Not. Roy. Astron. Soc.} {\bf 428} (2013) 1116--1127},
  [\href{http://arxiv.org/abs/1208.1491}{{\tt 1208.1491}}].

\bibitem{Leistedt:2014zqa}
B.~Leistedt, H.~V. Peiris and N.~Roth, \emph{{Constraints on Primordial
  Non-Gaussianity from 800 000 Photometric Quasars}},
  \href{http://dx.doi.org/10.1103/PhysRevLett.113.221301}{\emph{Phys. Rev.
  Lett.} {\bf 113} (2014) 221301}, [\href{http://arxiv.org/abs/1405.4315}{{\tt
  1405.4315}}].

\bibitem{Ho:2013lda}
S.~Ho et~al., \emph{{Sloan Digital Sky Survey III photometric quasar
  clustering: probing the initial conditions of the Universe}},
  \href{http://dx.doi.org/10.1088/1475-7516/2015/05/040}{\emph{JCAP} {\bf 05}
  (2015) 040}, [\href{http://arxiv.org/abs/1311.2597}{{\tt 1311.2597}}].

\bibitem{Castorina:2019wmr}
E.~Castorina et~al., \emph{{Redshift-weighted constraints on primordial
  non-Gaussianity from the clustering of the eBOSS DR14 quasars in Fourier
  space}}, \href{http://dx.doi.org/10.1088/1475-7516/2019/09/010}{\emph{JCAP}
  {\bf 09} (2019) 010}, [\href{http://arxiv.org/abs/1904.08859}{{\tt
  1904.08859}}].

\bibitem{Mueller:2021tqa}
E.-M. Mueller et~al., \emph{{The clustering of galaxies in the completed
  SDSS-IV extended Baryon Oscillation Spectroscopic Survey: Primordial
  non-Gaussianity in Fourier Space}},
  \href{http://arxiv.org/abs/2106.13725}{{\tt 2106.13725}}.

\bibitem{Cabass:2022wjy}
G.~Cabass, M.~M. Ivanov, O.~H.~E. Philcox, M.~Simonovi\'c and M.~Zaldarriaga,
  \emph{{Constraints on Single-Field Inflation from the BOSS Galaxy Survey}},
  \href{http://dx.doi.org/10.1103/PhysRevLett.129.021301}{\emph{Phys. Rev.
  Lett.} {\bf 129} (2022) 021301}, [\href{http://arxiv.org/abs/2201.07238}{{\tt
  2201.07238}}].

\bibitem{DAmico:2022gki}
G.~D'Amico, M.~Lewandowski, L.~Senatore and P.~Zhang, \emph{{Limits on
  primordial non-Gaussianities from BOSS galaxy-clustering data}},
  \href{http://arxiv.org/abs/2201.11518}{{\tt 2201.11518}}.

\bibitem{Cabass:2022ymb}
G.~Cabass, M.~M. Ivanov, O.~H.~E. Philcox, M.~Simonovi\'c and M.~Zaldarriaga,
  \emph{{Constraints on multifield inflation from the BOSS galaxy survey}},
  \href{http://dx.doi.org/10.1103/PhysRevD.106.043506}{\emph{Phys. Rev. D} {\bf
  106} (2022) 043506}, [\href{http://arxiv.org/abs/2204.01781}{{\tt
  2204.01781}}].

\bibitem{Chan:2016ehg}
K.~C. Chan and L.~Blot, \emph{{Assessment of the Information Content of the
  Power Spectrum and Bispectrum}},
  \href{http://dx.doi.org/10.1103/PhysRevD.96.023528}{\emph{Phys. Rev. D} {\bf
  96} (2017) 023528}, [\href{http://arxiv.org/abs/1610.06585}{{\tt
  1610.06585}}].

\bibitem{Biagetti:2021tua}
M.~Biagetti, L.~Castiblanco, J.~Nore\~na and E.~Sefusatti, \emph{{The
  covariance of squeezed bispectrum configurations}},
  \href{http://dx.doi.org/10.1088/1475-7516/2022/09/009}{\emph{JCAP} {\bf 09}
  (2022) 009}, [\href{http://arxiv.org/abs/2111.05887}{{\tt 2111.05887}}].

\bibitem{Coulton:2022qbc}
W.~R. Coulton, F.~Villaescusa-Navarro, D.~Jamieson, M.~Baldi, G.~Jung,
  D.~Karagiannis et~al., \emph{{Quijote-PNG: Simulations of Primordial
  Non-Gaussianity and the Information Content of the Matter Field Power
  Spectrum and Bispectrum}},
  \href{http://dx.doi.org/10.3847/1538-4357/aca8a7}{\emph{Astrophys. J.} {\bf
  943} (2023) 64}, [\href{http://arxiv.org/abs/2206.01619}{{\tt 2206.01619}}].

\bibitem{Floss:2022wkq}
T.~Fl\"oss, M.~Biagetti and P.~D. Meerburg, \emph{{Primordial non-Gaussianity
  and non-Gaussian covariance}},
  \href{http://dx.doi.org/10.1103/PhysRevD.107.023528}{\emph{Phys. Rev. D} {\bf
  107} (2023) 023528}, [\href{http://arxiv.org/abs/2206.10458}{{\tt
  2206.10458}}].

\bibitem{Goldstein:2022hgr}
S.~Goldstein, A.~Esposito, O.~H.~E. Philcox, L.~Hui, J.~C. Hill, R.~Scoccimarro
  et~al., \emph{{Squeezing fNL out of the matter bispectrum with consistency
  relations}}, \href{http://dx.doi.org/10.1103/PhysRevD.106.123525}{\emph{Phys.
  Rev. D} {\bf 106} (2022) 123525},
  [\href{http://arxiv.org/abs/2209.06228}{{\tt 2209.06228}}].

\bibitem{Eisenstein:2006nk}
D.~J. Eisenstein, H.-j. Seo, E.~Sirko and D.~Spergel, \emph{{Improving
  Cosmological Distance Measurements by Reconstruction of the Baryon Acoustic
  Peak}}, \href{http://dx.doi.org/10.1086/518712}{\emph{Astrophys. J.} {\bf
  664} (2007) 675--679}, [\href{http://arxiv.org/abs/astro-ph/0604362}{{\tt
  astro-ph/0604362}}].

\bibitem{Wang:2022nlx}
Y.~Wang et~al., \emph{{Extracting high-order cosmological information in galaxy
  surveys with power spectra}},  \href{http://arxiv.org/abs/2202.05248}{{\tt
  2202.05248}}.

\bibitem{Shirasaki:2020vkk}
M.~Shirasaki, N.~S. Sugiyama, R.~Takahashi and F.-S. Kitaura,
  \emph{{Constraining primordial non-Gaussianity with postreconstructed galaxy
  bispectrum in redshift space}},
  \href{http://dx.doi.org/10.1103/PhysRevD.103.023506}{\emph{Phys. Rev. D} {\bf
  103} (2021) 023506}, [\href{http://arxiv.org/abs/2010.04567}{{\tt
  2010.04567}}].

\bibitem{He:2018ggn}
S.~He, Y.~Li, Y.~Feng, S.~Ho, S.~Ravanbakhsh, W.~Chen et~al., \emph{{Learning
  to Predict the Cosmological Structure Formation}},
  \href{http://dx.doi.org/10.1073/pnas.1821458116}{\emph{Proc. Nat. Acad. Sci.}
  {\bf 116} (2019) 13825--13832}, [\href{http://arxiv.org/abs/1811.06533}{{\tt
  1811.06533}}].

\bibitem{Jamieson:2022lqc}
D.~Jamieson, Y.~Li, R.~A. de~Oliveira, F.~Villaescusa-Navarro, S.~Ho and D.~N.
  Spergel, \emph{{Field Level Neural Network Emulator for Cosmological N-body
  Simulations}},  \href{http://arxiv.org/abs/2206.04594}{{\tt 2206.04594}}.

\bibitem{Jamieson:2022daw}
D.~Jamieson, Y.~Li, S.~He, F.~Villaescusa-Navarro, S.~Ho, R.~A. de~Oliveira
  et~al., \emph{{Simple lessons from complex learning: what a neural network
  model learns about cosmic structure formation}},
  \href{http://arxiv.org/abs/2206.04573}{{\tt 2206.04573}}.

\bibitem{Jindal:2023qew}
V.~Jindal, D.~Jamieson, A.~Liang, A.~Singh and S.~Ho, \emph{{Predicting the
  Initial Conditions of the Universe using Deep Learning}},
  \href{http://arxiv.org/abs/2303.13056}{{\tt 2303.13056}}.

\bibitem{Shallue:2022mhf}
C.~J. Shallue and D.~J. Eisenstein, \emph{{Reconstructing cosmological initial
  conditions from late-time structure with convolutional neural networks}},
  \href{http://dx.doi.org/10.1093/mnras/stad528}{\emph{Mon. Not. Roy. Astron.
  Soc.} {\bf 520} (2023) 6256--6267},
  [\href{http://arxiv.org/abs/2207.12511}{{\tt 2207.12511}}].

\bibitem{Schmittfull:2017uhh}
M.~Schmittfull, T.~Baldauf and M.~Zaldarriaga, \emph{{Iterative initial
  condition reconstruction}},
  \href{http://dx.doi.org/10.1103/PhysRevD.96.023505}{\emph{Phys. Rev. D} {\bf
  96} (2017) 023505}, [\href{http://arxiv.org/abs/1704.06634}{{\tt
  1704.06634}}].

\bibitem{Villaescusa-Navarro:2019bje}
F.~Villaescusa-Navarro et~al., \emph{{The Quijote simulations}},
  \href{http://dx.doi.org/10.3847/1538-4365/ab9d82}{\emph{Astrophys. J. Suppl.}
  {\bf 250} (2020) 2}, [\href{http://arxiv.org/abs/1909.05273}{{\tt
  1909.05273}}].

\bibitem{Biagetti:2022ckz}
M.~Biagetti, J.~Calles, L.~Castiblanco, K.~Gonz\'alez and J.~Nore\~na, \emph{{A
  Model for the Squeezed Bispectrum in the Non-Linear Regime}},
  \href{http://arxiv.org/abs/2212.11940}{{\tt 2212.11940}}.

\bibitem{Babich:2004gb}
D.~Babich, P.~Creminelli and M.~Zaldarriaga, \emph{{The Shape of
  non-Gaussianities}},
  \href{http://dx.doi.org/10.1088/1475-7516/2004/08/009}{\emph{JCAP} {\bf 08}
  (2004) 009}, [\href{http://arxiv.org/abs/astro-ph/0405356}{{\tt
  astro-ph/0405356}}].

\bibitem{Maldacena:2002vr}
J.~M. Maldacena, \emph{{Non-Gaussian features of primordial fluctuations in
  single field inflationary models}},
  \href{http://dx.doi.org/10.1088/1126-6708/2003/05/013}{\emph{JHEP} {\bf 05}
  (2003) 013}, [\href{http://arxiv.org/abs/astro-ph/0210603}{{\tt
  astro-ph/0210603}}].

\bibitem{Creminelli:2004yq}
P.~Creminelli and M.~Zaldarriaga, \emph{{Single field consistency relation for
  the 3-point function}},
  \href{http://dx.doi.org/10.1088/1475-7516/2004/10/006}{\emph{JCAP} {\bf 10}
  (2004) 006}, [\href{http://arxiv.org/abs/astro-ph/0407059}{{\tt
  astro-ph/0407059}}].

\bibitem{Cheung:2007st}
C.~Cheung, P.~Creminelli, A.~L. Fitzpatrick, J.~Kaplan and L.~Senatore,
  \emph{{The Effective Field Theory of Inflation}},
  \href{http://dx.doi.org/10.1088/1126-6708/2008/03/014}{\emph{JHEP} {\bf 03}
  (2008) 014}, [\href{http://arxiv.org/abs/0709.0293}{{\tt 0709.0293}}].

\bibitem{Senatore:2009gt}
L.~Senatore, K.~M. Smith and M.~Zaldarriaga, \emph{{Non-Gaussianities in Single
  Field Inflation and their Optimal Limits from the WMAP 5-year Data}},
  \href{http://dx.doi.org/10.1088/1475-7516/2010/01/028}{\emph{JCAP} {\bf 01}
  (2010) 028}, [\href{http://arxiv.org/abs/0905.3746}{{\tt 0905.3746}}].

\bibitem{Carrasco:2013mua}
J.~J.~M. Carrasco, S.~Foreman, D.~Green and L.~Senatore, \emph{{The Effective
  Field Theory of Large Scale Structures at Two Loops}},
  \href{http://dx.doi.org/10.1088/1475-7516/2014/07/057}{\emph{JCAP} {\bf 07}
  (2014) 057}, [\href{http://arxiv.org/abs/1310.0464}{{\tt 1310.0464}}].

\bibitem{Angulo:2014tfa}
R.~E. Angulo, S.~Foreman, M.~Schmittfull and L.~Senatore, \emph{{The One-Loop
  Matter Bispectrum in the Effective Field Theory of Large Scale Structures}},
  \href{http://dx.doi.org/10.1088/1475-7516/2015/10/039}{\emph{JCAP} {\bf 10}
  (2015) 039}, [\href{http://arxiv.org/abs/1406.4143}{{\tt 1406.4143}}].

\bibitem{Coulton:2019odk}
W.~R. Coulton, P.~D. Meerburg, D.~G. Baker, S.~Hotinli, A.~J. Duivenvoorden and
  A.~van Engelen, \emph{{Minimizing gravitational lensing contributions to the
  primordial bispectrum covariance}},
  \href{http://dx.doi.org/10.1103/PhysRevD.101.123504}{\emph{Phys. Rev. D} {\bf
  101} (2020) 123504}, [\href{http://arxiv.org/abs/1912.07619}{{\tt
  1912.07619}}].

\bibitem{Schmittfull:2015mja}
M.~Schmittfull, Y.~Feng, F.~Beutler, B.~Sherwin and M.~Y. Chu, \emph{{Eulerian
  BAO Reconstructions and N-Point Statistics}},
  \href{http://dx.doi.org/10.1103/PhysRevD.92.123522}{\emph{Phys. Rev. D} {\bf
  92} (2015) 123522}, [\href{http://arxiv.org/abs/1508.06972}{{\tt
  1508.06972}}].

\bibitem{ronneberger2015u}
O.~Ronneberger, P.~Fischer and T.~Brox, \emph{U-net: Convolutional networks for
  biomedical image segmentation},  in \emph{Medical Image Computing and
  Computer-Assisted Intervention--MICCAI 2015: 18th International Conference,
  Munich, Germany, October 5-9, 2015, Proceedings, Part III 18}, pp.~234--241,
  Springer, 2015.

\bibitem{Makinen:2020gvh}
T.~L. Makinen, L.~Lancaster, F.~Villaescusa-Navarro, P.~Melchior, S.~Ho,
  L.~Perreault-Levasseur et~al., \emph{{deep21: a deep learning method for 21
  cm foreground removal}},
  \href{http://dx.doi.org/10.1088/1475-7516/2021/04/081}{\emph{JCAP} {\bf 04}
  (2021) 081}, [\href{http://arxiv.org/abs/2010.15843}{{\tt 2010.15843}}].

\bibitem{Gagnon-Hartman:2021erd}
S.~Gagnon-Hartman, Y.~Cui, A.~Liu and S.~Ravanbakhsh, \emph{{Recovering the
  Wedge Modes Lost to 21-cm Foregrounds}},
  \href{http://dx.doi.org/10.1093/mnras/stab1158}{\emph{Mon. Not. Roy. Astron.
  Soc.} {\bf 504} (2021) 4716}, [\href{http://arxiv.org/abs/2102.08382}{{\tt
  2102.08382}}].

\bibitem{Taruya:2018jtk}
A.~Taruya, T.~Nishimichi and D.~Jeong, \emph{{Grid-based calculation for
  perturbation theory of large-scale structure}},
  \href{http://dx.doi.org/10.1103/PhysRevD.98.103532}{\emph{Phys. Rev. D} {\bf
  98} (2018) 103532}, [\href{http://arxiv.org/abs/1807.04215}{{\tt
  1807.04215}}].

\bibitem{Sefusatti:2015aex}
E.~Sefusatti, M.~Crocce, R.~Scoccimarro and H.~Couchman, \emph{{Accurate
  Estimators of Correlation Functions in Fourier Space}},
  \href{http://dx.doi.org/10.1093/mnras/stw1229}{\emph{Mon. Not. Roy. Astron.
  Soc.} {\bf 460} (2016) 3624--3636},
  [\href{http://arxiv.org/abs/1512.07295}{{\tt 1512.07295}}].

\bibitem{Chen:2023uup}
X.~Chen, F.~Zhu, S.~Gaines and N.~Padmanabhan, \emph{{Effective cosmic density
  field reconstruction with convolutional neural network}},
  \href{http://dx.doi.org/10.1093/mnras/stad1868}{\emph{Mon. Not. Roy. Astron.
  Soc.} {\bf 523} (2023) 6272--6281},
  [\href{http://arxiv.org/abs/2306.10538}{{\tt 2306.10538}}].

\bibitem{Jing:2004fq}
Y.~P. Jing, \emph{{Correcting for the alias effect when measuring the power
  spectrum using FFT}},
  \href{http://dx.doi.org/10.1086/427087}{\emph{Astrophys. J.} {\bf 620} (2005)
  559--563}, [\href{http://arxiv.org/abs/astro-ph/0409240}{{\tt
  astro-ph/0409240}}].

\bibitem{Scoccimarro:2000sn}
R.~Scoccimarro, \emph{{The bispectrum: from theory to observations}},
  \href{http://dx.doi.org/10.1086/317248}{\emph{Astrophys. J.} {\bf 544} (2000)
  597}, [\href{http://arxiv.org/abs/astro-ph/0004086}{{\tt astro-ph/0004086}}].

\bibitem{Hartlap:2006kj}
J.~Hartlap, P.~Simon and P.~Schneider, \emph{{Why your model parameter
  confidences might be too optimistic: Unbiased estimation of the inverse
  covariance matrix}},
  \href{http://dx.doi.org/10.1051/0004-6361:20066170}{\emph{Astron. Astrophys.}
  {\bf 464} (2007) 399}, [\href{http://arxiv.org/abs/astro-ph/0608064}{{\tt
  astro-ph/0608064}}].

\bibitem{Valogiannis:2021chp}
G.~Valogiannis and C.~Dvorkin, \emph{{Towards an optimal estimation of
  cosmological parameters with the wavelet scattering transform}},
  \href{http://dx.doi.org/10.1103/PhysRevD.105.103534}{\emph{Phys. Rev. D} {\bf
  105} (2022) 103534}, [\href{http://arxiv.org/abs/2108.07821}{{\tt
  2108.07821}}].

\bibitem{Jung:2022rtn}
G.~Jung, D.~Karagiannis, M.~Liguori, M.~Baldi, W.~R. Coulton, D.~Jamieson
  et~al., \emph{{Quijote-PNG: Quasi-maximum Likelihood Estimation of Primordial
  Non-Gaussianity in the Nonlinear Dark Matter Density Field}},
  \href{http://dx.doi.org/10.3847/1538-4357/ac9837}{\emph{Astrophys. J.} {\bf
  940} (2022) 71}, [\href{http://arxiv.org/abs/2206.01624}{{\tt 2206.01624}}].

\end{thebibliography}\endgroup

\newpage
\appendix
\section{Information content of power spectrum and bispectrum separately}
\label{app:PandB}
\begin{figure}[H]
    \centering
    \includegraphics[scale=.70]{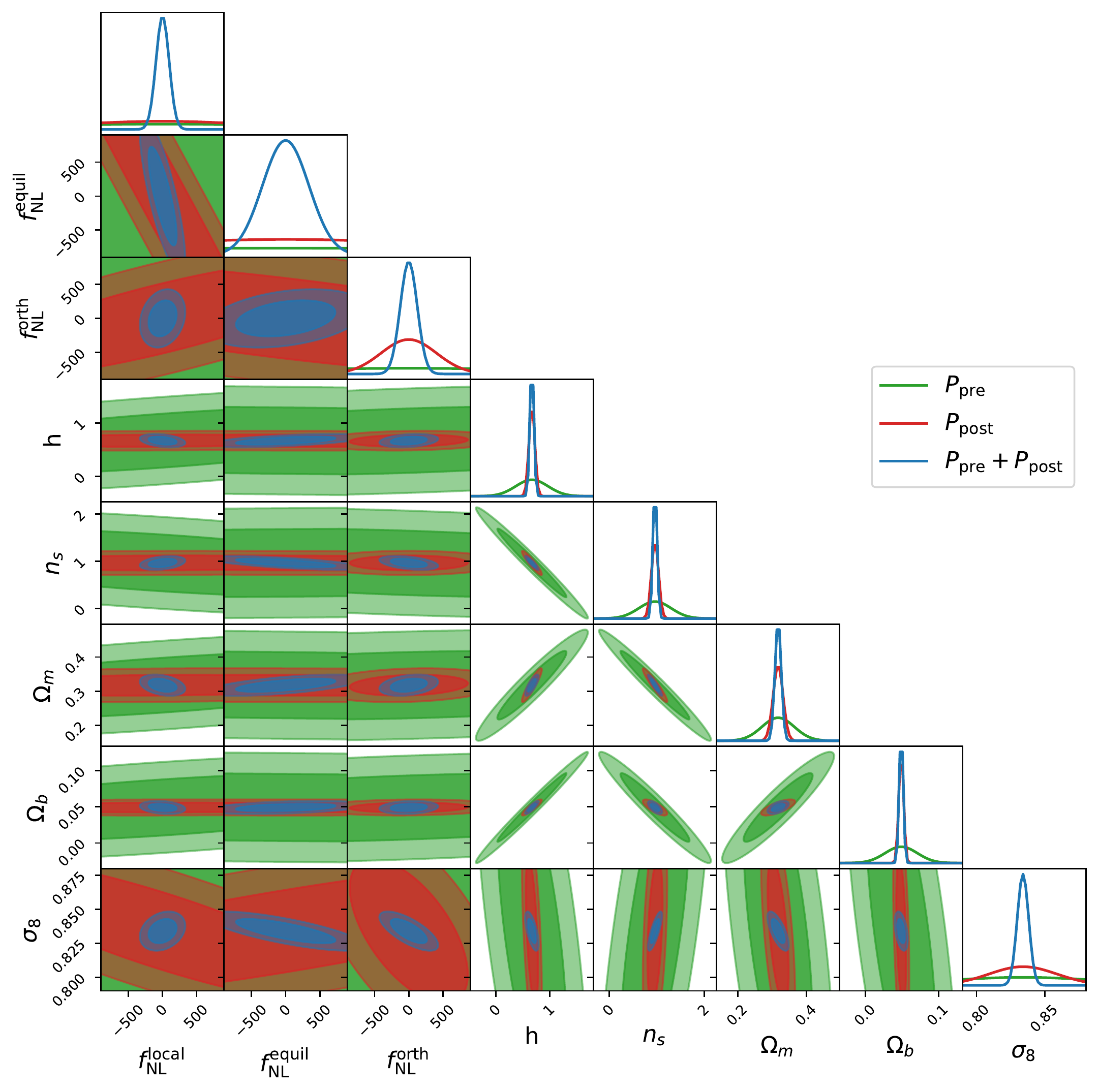}
    \caption{1 and 2-$\sigma$ marginalized constraints when jointly fitting for all eight parameters using the \textcolor{tabgreen}{pre}, \textcolor{tabred}{post} and \textcolor{tabblue}{pre+post} reconstructed power spectrum. Note that the constraints from only the pre-reconstructed power spectrum are likely too optimistic since the Fisher derivatives have not fully converged, as discussed in Appendix \ref{app:convergence} and Ref.~\cite{Coulton:2022qbc}.}
    \label{fig:confidenceplotP}
\end{figure}

\begin{figure}[H]
    \centering
    \includegraphics[scale=.70]{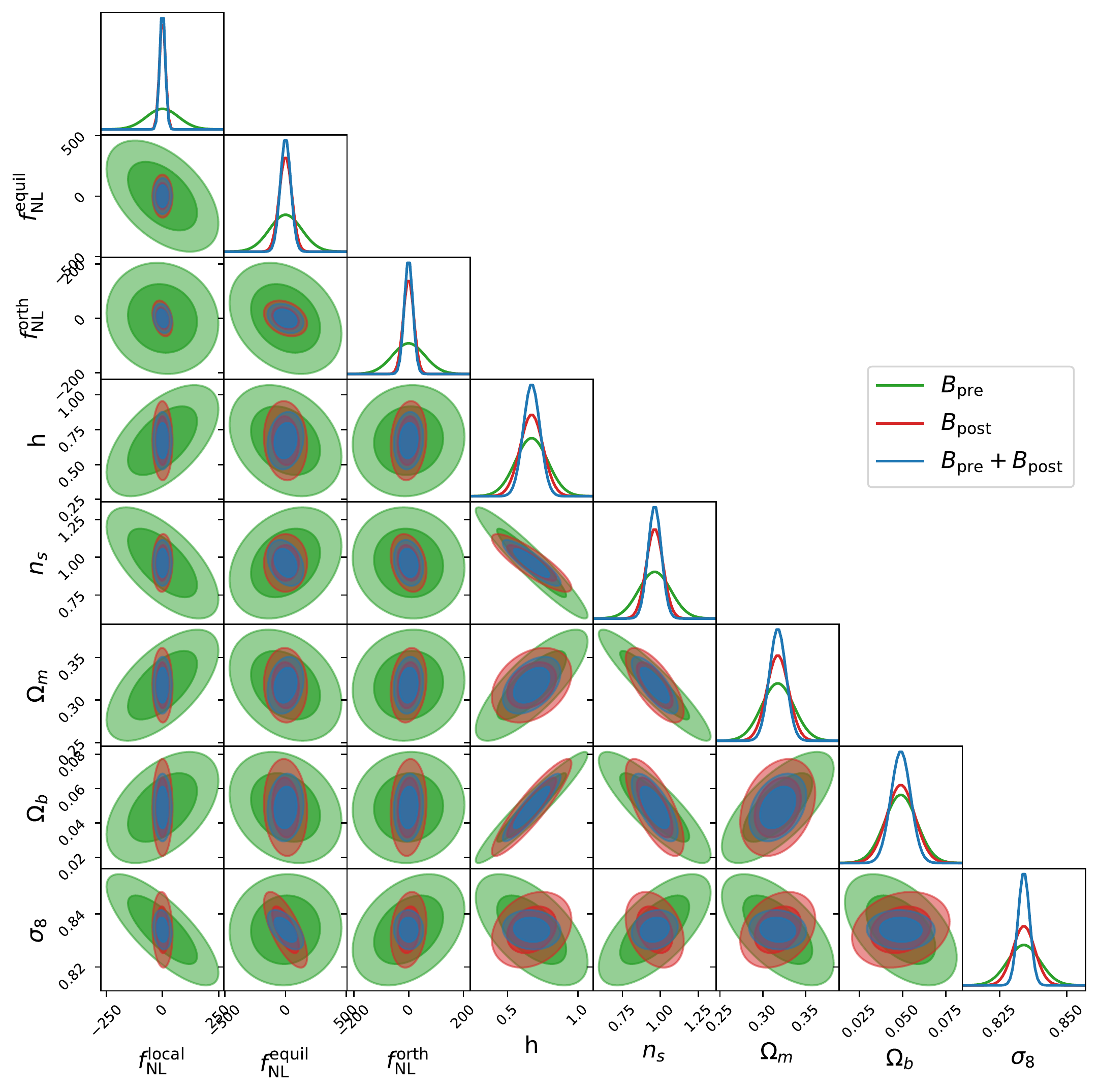}
    \caption{1 and 2-$\sigma$ marginalized constraints when jointly fitting for all eight parameters using the \textcolor{tabgreen}{pre}, \textcolor{tabred}{post} and \textcolor{tabblue}{pre+post} reconstructed bispectrum.}
    \label{fig:confidenceplotB}
\end{figure}

For completeness, we present in Figure \ref{fig:confidenceplotP} and Figure \ref{fig:confidenceplotB} the confidence intervals when using only the power spectrum or the bispectrum, respectively. By combining the pre and post-reconstructed power spectrum we gain significant information over either one of them alone, as was pointed out also in Ref.~\cite{Wang:2022nlx}.

\section{Convergence of Fisher analysis}
\label{app:convergence}
\begin{figure}
    \centering
    \includegraphics[scale=.53]{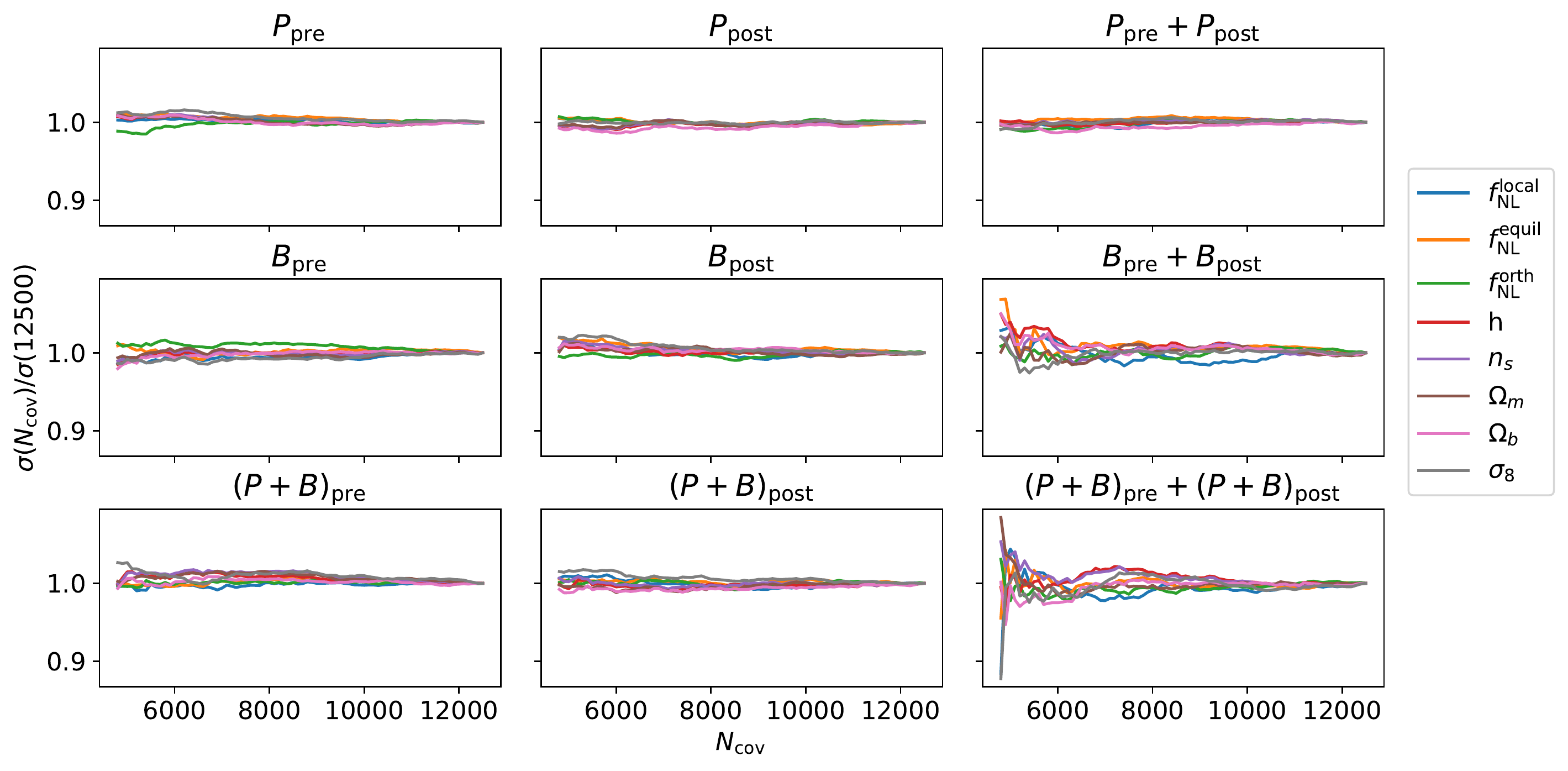}
    \caption{Variation of the marginalized parameter constraints as a function of the number of simulations included in the computation of the covariance matrix, for all the different data products. For this test, we keep the number of simulations used to compute derivatives fixed at 500.}
    \label{fig:convergence_cov}
\end{figure}

\begin{figure}
    \centering
    \includegraphics[scale=.53]{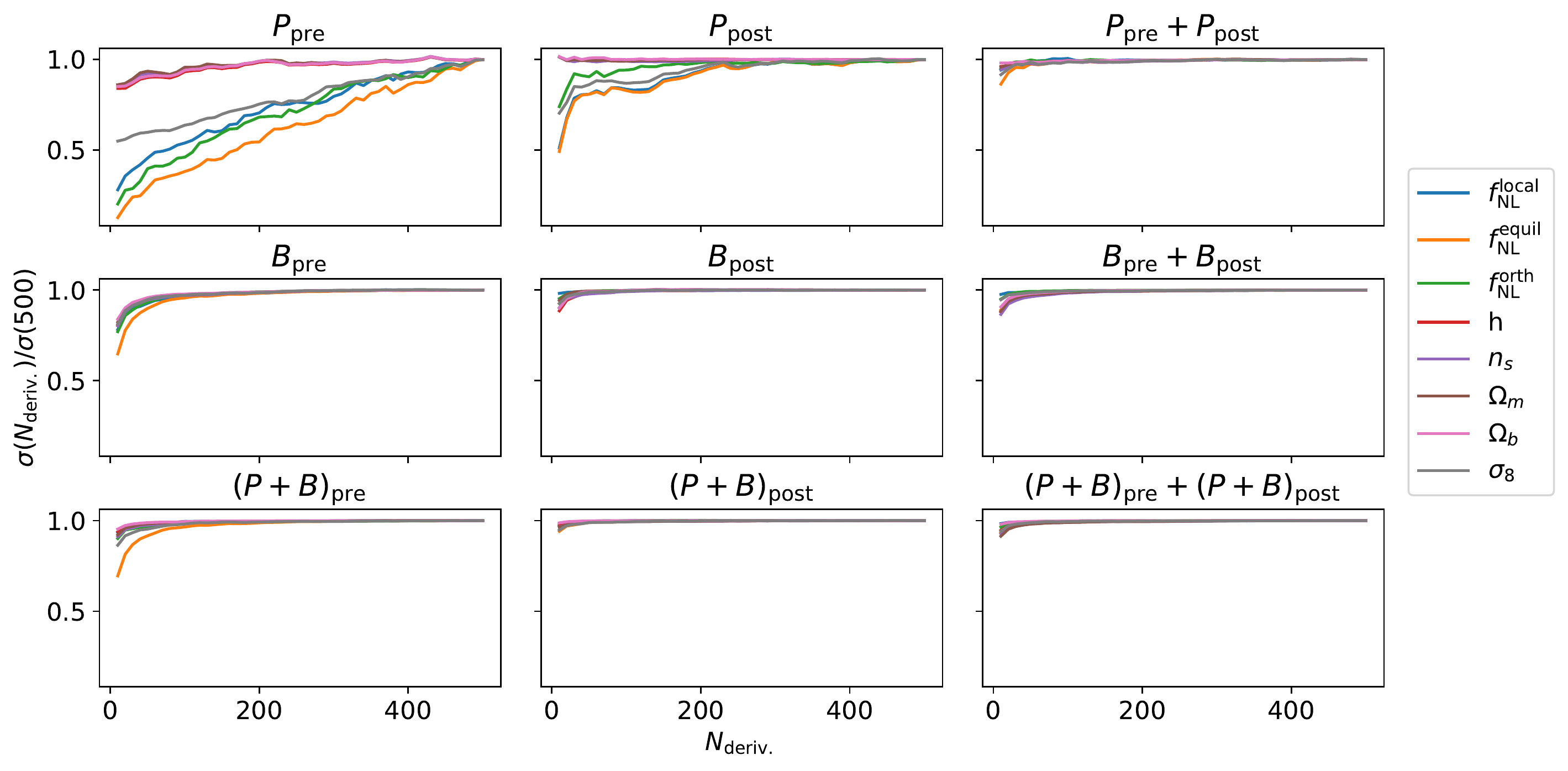}
    \caption{Variation of the marginalized parameter constraints as a function of the number of simulations included in the computation of the Fisher derivatives, for all the different data products. For this test, we keep the number of simulations used to compute the covariance matrix fixed at 12500.}
    \label{fig:convergence_deriv}
\end{figure}

We present here tests of the convergence of the Fisher forecasts presented in the main text. First, we assess whether the covariance matrix has converged sufficiently to provide robust parameter constraints. In Figure \ref{fig:convergence_cov}, we plot the variation of the marginalized parameter constraints from a Fisher analysis, as a function of the number of simulations included in the computation of the covariance matrix, while keeping the number of simulations used to compute derivatives fixed at maximum (i.e. 500). As we can see, the parameter constraints from all data products are increasingly stable as we include more simulations and appear to have converged well at 12500 simulations. We thus conclude that we have used a sufficient number of simulations to compute the covariance matrix in our analysis. Next, we test whether the derivatives that enter the Fisher matrix computation of equation \eqref{eq:fish}, have converged. In Figure \ref{fig:convergence_deriv}, we show the variation of the marginalized parameters constraints, as a function of the number of simulations included in the computation of the derivatives, while keeping the number of simulations used to compute the covariance matrix fixed at maximum (i.e. 12500). As was found in the original work of Ref.~\cite{Coulton:2022qbc}, the pre-reconstructed power spectrum has not converged and hence the forecasts based on this data alone should not be trusted. However, the parameter constraints for all the other data products seem to have converged well when using 500 simulations to compute the derivatives.

\end{document}